\documentclass[journal,11pt]{IEEEtran}
 
\usepackage{cite}
\usepackage[dvips]{graphicx}
\usepackage[cmex10]{amsmath}
\usepackage{color}
\usepackage{bm}
\usepackage{theorem} 
\usepackage{amscd}  
\usepackage{amssymb}
\usepackage {subfigmat}  
\usepackage{multirow}    
\usepackage{mathtools}       
                 
{\theorembodyfont{\normalfont\it}
\theoremheaderfont{\normalfont\bf}
\newtheorem{thm}{ Theorem}
\newtheorem{dfn}[thm]{ Definition}
\newtheorem{lmm}[thm]{ Lemma}

\newtheorem{crl}[thm]{ Corollary}
\newtheorem{asm}[thm]{ Assumption}
\newtheorem{prp}[thm]{ Proposition}
\newtheorem{cjt}[thm]{ Conjecture}}

{\theorembodyfont{\normalfont}
\theoremheaderfont{\normalfont\it}
\newtheorem{prf}{ Proof:}}

{\theorembodyfont{\normalfont}
\theoremheaderfont{\normalfont\it}
}

{\theorembodyfont{\normalfont}
\theoremheaderfont{\normalfont\it}
}

{\theorembodyfont{\normalfont}
\theoremheaderfont{\normalfont\it}
}

{\theorembodyfont{\normalfont}
\theoremheaderfont{\normalfont\it}
}

{\theorembodyfont{\normalfont}
\theoremheaderfont{\normalfont\it}
}

{\theorembodyfont{\normalfont}
\theoremheaderfont{\normalfont\it}
\newtheorem{rmk}{ Remark.}}

\newcommand{\bra}[1]{\mbox{$\langle#1|$}}

\newcommand{\ket}[1]{\mbox{$|#1\rangle$}}

\newcommand{\proj}[1]{\mbox{$\ket{#1}\!\bra{#1}$}}

\newcommand{\alg}[1]{\begin{align}#1\end{align}}

\newcommand{\nn}{\nonumber}
\newcommand{\ca}[1]{{\mathcal #1}}
\newcommand{\mbb}[1]{{\mathbb #1}}
\newcommand{\mfk}[1]{{\mathfrak #1}}

\newcommand{\bthm}[1]{\begin{thm}\label{thm:#1}}
\newcommand{\ethm}{\end{thm}}

\newcommand{\rThm}[1]{Theorem \ref{thm:#1}}
\newcommand{\blmm}[1]{\begin{lmm}\label{lmm:#1}}
\newcommand{\elmm}{\end{lmm}}
\newcommand{\rLmm}[1]{Lemma \ref{lmm:#1}}
\newcommand{\rlmm}[1]{\ref{lmm:#1}}
\newcommand{\bdfn}[1]{\begin{dfn}\label{dfn:#1}}
\newcommand{\edfn}{\end{dfn}}

\newcommand{\basm}[1]{\begin{asm}\label{asm:#1}}
\newcommand{\easm}{\end{asm}}
\newcommand{\bprp}[1]{\begin{prp}\label{prp:#1}}
\newcommand{\eprp}{\end{prp}}

\newcommand{\rPrp}[1]{Proposition \ref{prp:#1}}
\newcommand{\bcrl}[1]{\begin{crl}\label{crl:#1}}
\newcommand{\ecrl}{\end{crl}}

\newcommand{\rCrl}[1]{Corollary \ref{crl:#1}}
\newcommand{\bcjt}[1]{\begin{cjt}\label{cjt:#1}}
\newcommand{\ecjt}{\end{cjt}}

\newcommand{\bprf}{\begin{prf}}
\newcommand{\eprf}{\end{prf}}
\newcommand{\brmk}{\begin{rmk}}
\newcommand{\ermk}{\end{rmk}}
\newcommand{\laeq}[1]{\label{eq:#1}}
\newcommand{\req}[1]{(\ref{eq:#1})}

\newcommand{\QED}{\hfill$\blacksquare$}
\newcommand{\lsec}[1]{\label{sec:#1}}
\newcommand{\rsec}[1]{\ref{sec:#1}}
\newcommand{\rSec}[1]{Section \ref{sec:#1}}

\newcommand{\bitem}{\begin{itemize}}
\newcommand{\entem}{\end{itemize}}
\newcommand{\benum}{\begin{enumerate}}
\newcommand{\ennum}{\end{enumerate}}
\newcommand{\bb}{\mathbb}
\newcommand{\otm}{\otimes}

\newcommand{\beq}{\begin{eqnarray}}
\newcommand{\eeq}{\end{eqnarray}}

\begin{document}

\title{Quantum Multiple-Access One-Time Pad}

\author{Eyuri Wakakuwa

\thanks{E. Wakakuwa is with the Department of Computer Science, Graduate School of Information Science and Technology,
The University of Tokyo, 7-3-1, Hongo, Bunkyo-ku, Tokyo, 113-0033, Japan (email: e.wakakuwa@gmail.com).}
}

\maketitle

\begin{abstract}
We introduce and analyze an information theoretical task that we call the {\it quantum multiple-access one-time pad}.
Here, a number of senders initially share a correlated quantum state with a receiver and an eavesdropper.
Each sender performs a local operation to encode a classical message and sends their system to the receiver, who subsequently performs a measurement to decode the messages.
The receiver will be able to decode the messages almost perfectly, while the eavesdropper must not be able to extract information about the messages even if they have access to the quantum systems transmitted.
We consider a ``conditional'' scenario in which a portion of the receiver's side information is also accessible to the eavesdropper.
We investigate the maximum amount of classical information that can be encoded by each of the senders.
We derive a single-letter characterization for the achievable rate region in an asymptotic limit of infinitely many copies and vanishingly small error.
\end{abstract}


\section{Introduction}

Encoding and decoding of classical information into/from a composite quantum system under locality restrictions have been one of the central issues in quantum information theory.
Superdense coding \cite{bennett1992communication} is a task of encoding classical information to a bipartite quantum state by local operations on one of the two subsystems.
Quantum secret sharing \cite{hillery1999quantum,karlsson1999quantum} and quantum data hiding \cite{terhal2001hiding,divincenzo2001quantum,eggeling2002hiding,divincenzo2003hiding} are schemes that prevent anyone under locality restrictions from extracting the information encoded in a multipartite quantum system.
The quantum one-time pad \cite{schumacher2006quantum,brandao2012quantum} and the conditional quantum one-time pad \cite{sharma2020conditional} are hybrid of the two scenarios, in the sense that classical information is encoded by local operations on one of the subsystems so that it can be decoded only if the global access to the entire system is granted.
These studies have lead to a better understanding of correlation and entanglement in multipartite quantum states from an operational and information theoretical viewpoint.

This paper considers an extension of the conditional quantum one-time pad to a multi-sender scenario.
We refer the task as the {\it quantum multiple-access one-time pad}.
Here, a receiver, an eavesdropper and a number of senders initially share a correlated quantum state.
Each sender performs a local operation to encode a classical message on that state.
The senders then send their systems to the receiver, who subsequently performs a measurement to decode the messages.
We require that the receiver can decode the messages almost perfectly, while the eavesdropper can obtain almost no information about the messages even if they have access to the quantum systems transmitted.
We focus on a ``conditional'' scenario as in \cite{sharma2020conditional}, i.e., we assume that a portion of the receiver's side information is accessible to the eavesdropper.
We consider an asymptotic limit of infinitely many copies and vanishingly small error,
and investigate the maximum amount of classical information that can be encoded by each of the senders.
The main result is that we derive a single-letter characterization for the achievable rate region.

This paper is organized as follows.
In \rSec{formandres}, we present the formulation of the problem and describe the main result. 
\rSec{DEDR} introduces two subprotocols that we call {\it distributed encoding} and {\it distributed randomization}, and present the coding theorems thereof.
Based on these results, we prove the main result in \rSec{prfmain}.
The proofs of the coding theorems for distributed encoding and distributed randomization are provided in \rSec{DistEnc} and \rSec{DistRand}, respectively,
and proofs of two lemmas that are used therein will be provided in \rSec{PRFtrans}.
Conclusions are given in \rSec{cncl}.\\

\noindent
{\bf Notations:}
For a natural number $N\in\mbb{N}$, the set of natural numbers no greater than $N$ is denoted by $[N]$, i.e., $[N]\equiv\{1,\cdots,N\}$.
The set of linear operators, unitary operators, normalized density operators and subnormalized ones on a Hilbert space $\ca{H}$ are denoted by $\ca{L}(\ca{H})$, $\ca{U}(\ca{H})$, $\ca{S}(\ca{H})$ and $\ca{S}_\leq(\ca{H})$, respectively.
The Hilbert space associated with a system $A$ is denoted by ${\mathcal H}^A$, and its dimension is denoted by $d_A$. 
The identity operator on $\ca{H}^A$ is denoted by $I^A$,
and the completely mixed state on system $A$ is denoted by $\pi^A$, i.e., $\pi^A=I^A/d_A$.
The system composed of two subsystems $A$ and $B$ is denoted by $AB$. 
The Hilbert space associated with a composite system $AB$ is denoted by ${\mathcal H}^{AB}$, i.e., ${\mathcal H}^{AB}={\mathcal H}^{A}\otm{\mathcal H}^{B}$.
When $M$  and $N$ are linear operators on ${\mathcal H}^A$ and ${\mathcal H}^B$, respectively, their tensor product $M\otimes N$ is denoted by $M^A\otimes N^B$ for clarity. 
The identity operation on system $A$ is denoted by ${\rm id}^A$.
When ${\mathcal E}$ is a quantum operation on $A$ and $\rho$ is a state on $AB$, the state $({\mathcal E}\otimes{\rm id}^B)(\rho^{AB})$ is denoted simply by
${\mathcal E}^A(\rho^{AB})$.  
For $\rho^{AB}$, $\rho^{A}$ represents ${\rm Tr}_B[\rho^{AB}]$. 
The system composed of $n$ identical systems of $A$ is denoted by $A^n$, and the corresponding Hilbert space is denoted by $({\mathcal H}^A)^{\otimes n}$ or ${\mathcal H}^{A^n}$.
The trace norm of an operator $X\in\ca{L}(\ca{H})$ is defined by $\|X\|_1:={\rm Tr}[\sqrt{X^\dagger X}]$ and the trace distance between two states $\rho,\sigma\in\ca{S}(\ca{H}^A)$ is defined by $\frac{1}{2}\|\rho-\sigma\|_1$.
 $\log{x}$ represents the base $2$ logarithm of $x$.
 The binary entropy is defined by $h(x):=-x\log{x}-(1-x)\log{(1-x)}$ and satisfies $\lim_{x\downarrow0}h(x)=0$.
 The cardinality of a set $S$ is denoted by $|S|$.

The von Neumann entropy of a state $\rho\in\ca{S}(\ca{H}^A)$ is defined by
\alg{
S(A)_\rho
=
S(\rho^A)
:=
-{\rm Tr}[\rho^{A}\log{\rho^{A}}].
}
For $\varrho\in\ca{S}(\ca{H}^{AB})$ and $\varsigma\in\ca{S}(\ca{H}^{ABC})$, the conditional entropy, the mutual information and the conditional mutual information are defined by
\alg{
S(A|B)_\varrho
&:=S(AB)_\varrho-S(B)_\varrho,
\\
I(A:B)_\varrho
&:=S(A)_\varrho-S(A|B)_\varrho,
\\
I(A:B|C)_\varsigma
&:=S(A|C)_\varsigma-S(A|BC)_\varsigma.
}
For the properties of the entropies and the mutual informations used in this paper, see e.g.~\cite{wildetext}.

\section{Formulation and Result}
\lsec{formandres}

Suppose that $Z\in\mbb{N}$ senders, a receiver and an eavesdropper are located distantly.
Let $A_1,\cdots,A_Z$, $B$ and $E$ be quantum systems in their possession, which are represented by finite-dimensional Hilbert spaces  $\ca{H}^{A_1},\cdots,\ca{H}^{A_Z}$, $\ca{H}^B$ and $\ca{H}^E$, respectively.
They initially share $n\in\mbb{N}$ copies of a quantum state $\rho\in\ca{S}(\ca{H}^{A_1\cdots A_ZBE})$.
Each sender encodes a classical message by performing an operation on their system locally.
For $z\in[Z]$, let $R_z$ be the bit length of the message that the $z$-th sender is to encode, divided by $n$, and let $M_z\equiv 2^{nR_z}$.
The encoding scheme is represented by a finite-dimensional quantum system $A_z'$ and a set of encoding quantum operations (completely positive trace-preserving maps) $\mfk{E}_z\equiv\{\ca{E}_{z,m_z}\}_{m_z=1}^{M_z}$ from $A_z^n$ to $A_z'$, where $m_z\in[M_z]$ denotes the values of the message.
For simplicity, we introduce the following notations:
\alg{
{\bf M}&\equiv [M_1]\times\cdots\times [M_Z],
\\
{\bf m}&\equiv (m_1,\cdots,m_Z).
}

After performing the encoding operations, the senders send their systems to the receiver, who subsequently performs a measurement to decode the messages.
We consider the ``conditional'' scenario of \cite{sharma2020conditional}, in which the receiver has access to $A_1'\cdots A_Z'B^nE^n$ when decoding the messages and the eavesdropper has access to $A_1'\cdots A_Z'E^n$ when trying to extract the information about the messages. 
The decoding measurement by the receiver is represented by a POVM $\mathfrak{M}\equiv\{\Lambda_{{\bf m}}\}_{{\bf m}\in {\bf M}}$ on $A_1'\cdots A_Z'B^nE^n$.
We refer to the tuple $\mfk{C}\equiv(\mfk{E}_1,\cdots,\mfk{E}_Z,\mathfrak{M})$ as a $(n,M_1,\cdots,M_Z)$ {\it code for} $\rho$.

For each message value ${\bf m}$, the state after the encoding operations is represented by
\alg{
\rho_{{\bf m}}
:=
\left(\bigotimes_{z=1}^Z\ca{E}_{z,m_z}^{A_z^n\rightarrow A_z'}\right)
(\rho^{\otimes n}).
\laeq{rhovecm0}
}
We assume that the messages are distributed uniformly, in which case the average state is
\alg{
\bar{\rho}
:=
\frac{1}{|{\bf M}|}
\sum_{{\bf m}\in {\bf M}}
\rho_{{\bf m}}.
\laeq{rhobar0}
}
The average decoding error is defined by
\alg{
\epsilon(\mfk{C})
:=
1-
\frac{1}{|{\bf M}|}
\sum_{{\bf m}\in {\bf M}}
{\rm Tr}[\rho_{{\bf m}}\Lambda_{{\bf m}}].
}
The average information leakage is quantified by the trace distance as
\alg{
\vartheta(\mfk{C})
:=
\frac{1}{|{\bf M}|}
\sum_{{\bf m}\in {\bf M}}
\left\|\rho_{{\bf m}}^{A_{[Z]}'E^n}-\bar{\rho}^{A_{[Z]}'E^n}\right\|_1.
\!\!
}
We require that both the decoding error and the information leakage can be made arbitrarily small in the asymptotic limit of $n$ to infinity.
A rigorous definition of the achievable rate region is as follows:

\bdfn{}
A rate tuple $(R_1,\cdots,R_Z)$ is achievable for the state $\rho\in\ca{S}(\ca{H}^{A_1\cdots A_ZBE})$ if, for any $\epsilon,\vartheta>0$ and any sufficiently large $n\in\mbb{N}$, there exists a $(n,2^{nR_1},\cdots,2^{nR_Z})$ code $\mfk{C}$ for $\rho$ that satisfies the reliability condition $\epsilon(\mfk{C})\leq\epsilon$ and the secrecy condition $\vartheta(\mfk{C})\leq\vartheta$.
The achievable rate region for $\rho$ is the closure of the set of all achievable rate tuples in $\mbb{R}^Z$.
\edfn

\noindent
The main result of this paper is the following theorem:

\bthm{mainthm}
The achievable rate region is equal to the set of all rate tuples $(R_1,\cdots,R_Z)\in\mbb{R}^Z$ that satisfy the following inequality for any $\Gamma\in[Z]$:
\alg{
\sum_{z\in\Gamma}R_z
\leq
I(A_\Gamma:A_{\Gamma_c}B|E)_\rho,
\laeq{maincond}
}
where $A_\Gamma$ and $A_{\Gamma_c}$ denote the systems composed of $\{A_z\}_{z\in\Gamma}$ and $\{A_z\}_{z\in\Gamma_c}$, respectively, with $\Gamma_c\equiv[Z]\backslash\Gamma$.
\ethm

\noindent
A proof of \rThm{mainthm} will be provided in \rSec{prfmain} based on the results presented in \rSec{DEDR}.

\section{Distributed Encoding and Distributed Randomization}
\lsec{DEDR}

We introduce two subprotocols that we call {\it distributed encoding} and {\it distributed randomization}, and present the coding theorems thereof.
We compare the results with those in the previous literature.
The two subprotocols are combined in \rSec{prfmain} to prove the direct part of \rThm{mainthm}.
The proofs of the two theorems will be provided in \rSec{DistEnc} and \rsec{DistRand}, respectively.

\subsection{Distributed Encoding}
\lsec{DEThm}

Suppose that $Z\in\mbb{N}$ senders and a receiver are located distantly.
Let $A_1,\cdots,A_Z$ and $V$ be quantum systems in their possession, which are represented by finite-dimensional Hilbert spaces  $\ca{H}^{A_1},\cdots,\ca{H}^{A_Z}$ and $\ca{H}^V$, respectively.
They initially share $n\in\mbb{N}$ copies of a quantum state $\rho\in\ca{S}(\ca{H}^{A_1\cdots A_ZV})$.
Distributed encoding is a task in which each sender encodes a classical message by performing a local unitary operation on $A_z^n$
 so that it can be decoded from $A_1^n\cdots A_Z^nV^n$ almost perfectly. 
For $z\in[Z]$, let $C_z$ be the bit length of the message that the $z$-th sender is to encode, divided by $n$, and let $K_z\equiv2^{nC_z}$.
We define ${\bf K}\equiv[K_1]\times\cdots\times[K_Z]$.
Each element of $\bf{K}$ is denoted as ${\bf k}\equiv(k_1,\cdots, k_Z)$, where $k_z\in[K_z]$.
A rigorous definition is as follows:

\bdfn{encode1}
For each $z\in[Z]$, let $\mfk{U}_z\equiv\{U_{z,k_z}\}_{k_z=1}^{K_z}$ be a set of unitaries on $A_z^n$.
For each ${\bf k}\in {\bf K}$, define a unitary
\alg{
U_{{\bf k}}:=\bigotimes_{z=1}^ZU_{z,k_z}
}
and a state
\alg{
\rho_{{\bf k}}
:=
(U_{{\bf k}}\otm I^{V^n})\rho^{\otimes n}(U_{{\bf k}}\otm I^{V^n})^\dagger.
\laeq{rhovecm}
}
A tuple $(\mfk{U}_1,\cdots,\mfk{U}_Z)$ is a $(n,K_1,\cdots,K_Z)$ distributed encoding for $\rho$ with error $\epsilon$ if there exists a measurement $\{\Lambda_{{\bf k}}\}_{{\bf k}\in {\bf K}}$ on $A_1^n\cdots A_Z^nV^n$ such that
\alg{
\frac{1}{|{\bf K}|}
\sum_{{\bf k}\in {\bf K}}
{\rm Tr}[\rho_{{\bf k}}\Lambda_{{\bf k}}]
\geq
1-\epsilon.
\!\!
\laeq{measerror}
}
\edfn

\noindent
Our interest is on how much classical information can be encoded in this manner.
We consider a scenario where the encoding unitaries are chosen independently and randomly according to the Haar measure.
A rigorous definition is as follows:

\bdfn{encode2}
A rate tuple $(C_1,\cdots,C_Z)$ is achievable in distributed encoding on a state $\rho\in\ca{S}(\ca{H}^{A_1\cdots A_ZV})$ if, for any $\epsilon,\xi>0$ and any sufficiently large $n\in\mbb{N}$, the following statement holds:

Let $U_{z,k_z,i}\in\ca{U}(\ca{H}^{A_z})$ for $i\in[n]$
and suppose that we choose $U_{z,k_z,i}$ randomly and independently according to the Haar measure for each $i\in[n]$, $k_z\in[2^{nC_z}]$ and $z\in[Z]$.
Let 
\alg{
U_{z,k}:=\bigotimes_{i=1}^nU_{z,k_z,i}
\laeq{encunitTP}
}
and $\mfk{U}_z\equiv\{U_{z,k_z}\}_{k_z=1}^{K_z}$.
Then, with probability no smaller than $1-\xi$, the tuple $(\mfk{U}_1,\cdots,\mfk{U}_Z)$ is a $(n, 2^{nC_1},\cdots,2^{nC_Z})$ distributed encoding for $\rho$ with error $\epsilon$.

The achievable rate region for $\rho$ is the closure of the set of all achievable rate tuples in $\mbb{R}^Z$.
\edfn

\noindent
A single-letter characterization of the achievable rate region is provided by the following proposition:

\bprp{iindayo1}
The achievable rate region for distributed encoding on $\rho$ is equal to the set of all tuples $(C_1,\cdots,C_Z)\in\mbb{R}^Z$ that satisfy
\alg{
\sum_{z\in\Gamma}C_z
\leq
\hat{C}(\Gamma)_\rho
\laeq{dontasp}
}
for any $\Gamma\in[Z]$, where
\alg{
 \hat{C}(\Gamma)_\rho:=\sum_{z\in\Gamma}\log{d_{A_z}}-S(A_\Gamma|A_{\Gamma_c}V)_\rho.
\laeq{dfnChat}
}
\eprp

\noindent
A proof of \rPrp{iindayo1} will be presented in \rSec{DistEnc}.
Although the converse part of \rPrp{iindayo1} is not necessary to prove the main theorem, we provide its proof for the completeness.
A few properties of $\hat{C}(\Gamma)_\rho$ are described in \rSec{submodularity} and will be used in the following sections to prove the main results.
We remark that the validity of the time sharing scheme used in the proof of the direct part of \rPrp{iindayo1} (see \rSec{DistEnc}) follows from the fact that the encoding unitaries are in the form of the tensor product as \req{encunitTP}.

\subsection{Distributed Randomization}
\lsec{DRThm}

Consider quantum systems $A_1,\cdots,A_Z$ and $W$ that are represented by finite-dimensional Hilbert spaces  $\ca{H}^{A_1},\cdots,\ca{H}^{A_Z}$ and $\ca{H}^W$, respectively.
Suppose that $n\in\mbb{N}$ copies of a state $\rho\in\ca{S}(\ca{H}^{A_1\cdots A_ZW})$ are distributed to $Z+1$ distant parties.
The task of distributed randomization is to transform the state by applying a random unitary operation individually and independently on each $A_z^n$, so that the state will become the product of the maximally mixed state on $A_1^n\cdots A_Z^n$ and a state on $W^n$. 
For $z\in[Z]$, let $L_z\equiv2^{nD_z}$ be the number of unitaries that are randomly applied on $A_z^n$.
A rigorous definition is as follows:

\bdfn{}
For each $z\in[Z]$, let $\mfk{U}_z\equiv\{U_{z,l_z}\}_{l_z=1}^{L_z}$ be a set of unitaries on $A_z^n$.
Consider random unitary operations $\ca{R}_z\:(z\in[Z])$ defined by
\alg{
\ca{R}_z(\cdot)
:=
\frac{1}{L_z}\sum_{l_z=1}^{L_z}U_{z,l_z}(\cdot)U_{z,l_z}^\dagger
\laeq{dfnRk}
}
and let
\alg{
\bar{\rho}
:=
\left(\bigotimes_{z=1}^Z\ca{R}_z\right)(\rho^{\otimes n}).
\laeq{randState}
}
A tuple $(\mfk{U}_1,\cdots,\mfk{U}_Z)$ is a $(n,L_1,\cdots,L_Z)$ distributed randomization of $\rho$ with error $\vartheta$ if 
it holds that
\alg{
\left\|\bar{\rho}^{A_{[Z]}^nW^n}
-
\pi^{A_{[Z]}^n}\otm(\rho^{\otm n})^{W^n}\right\|_1
\leq
\vartheta,
\!\!
\laeq{randCond}
}
where $\pi$ is the maximally mixed state on $A_{[Z]}^n$.
\edfn

\noindent
Our interest is on how much randomness is required to accomplish this task in an asymptotic limit of infinitely many copies and vanishingly small error.
We consider a scenario where each element of the set of the unitaries is chosen randomly and independently according to the Haar measure.
A rigorous definition is as follows:

\bdfn{}
A rate tuple $(D_1,\cdots,D_Z)$ is achievable in distributed randomization of a state $\rho\in\ca{S}(\ca{H}^{A_1\cdots A_ZW})$ if, for any $\vartheta,\xi>0$ and any sufficiently large $n\in\mbb{N}$, the following statement holds:

Let $U_{z,l_z,i}\in\ca{U}(\ca{H}^{A_z})$ for $i\in[n]$ and suppose that we choose $U_{z,l_z,i}$ randomly and independently according to the Haar measure for each $i\in[n]$, $l_z\in[2^{nD_z}]$ and $z\in[Z]$.
Let 
\alg{
U_{z,l}:=\bigotimes_{i=1}^nU_{z,l_z,i}
\laeq{randunitTP}
}
and $\mfk{U}_z\equiv\{U_{z,l}\}_{l=1}^{2^{nD_z}}$.
Then, with probability no smaller than $1-\xi$, the tuple $(\mfk{U}_1,\cdots,\mfk{U}_Z)$ is a $(n,2^{nD_1},\cdots,2^{nD_Z})$ distributed randomization of $\rho$ with error $\vartheta$.

The achievable rate region for distributed randomization of $\rho$ is the closure of the set of all achievable rate tuples in $\mbb{R}^Z$.
\edfn

\noindent
A single-letter characterization of the achievable rate region is provided by the following proposition:

\bprp{iindayo2}
The achievable rate region for distributed randomization of $\rho$ is equal to the set of all tuples $(D_1,\cdots,D_Z)\in\mbb{R}^Z$ that satisfy
\alg{
\sum_{z\in\Gamma}D_z
\geq
\hat{D}(\Gamma)_\rho
\laeq{dontask}
}
for any $\Gamma\in[Z]$, where
\alg{
\hat{D}(\Gamma)_\rho:=\sum_{z\in\Gamma}\log{d_{A_z}}-S(A_\Gamma|W)_\rho.
\laeq{dfnDhat}
}
\eprp

\noindent
A proof of \rPrp{iindayo2} will be given in \rSec{DistRand}.
Although the converse part of \rPrp{iindayo2} is not necessary to prove the main theorem, we provide its proof for the completeness.
A few properties of $\hat{D}(\Gamma)_\rho$ are described in \rSec{submodularity} and will be used in the following sections to prove the main results.
We remark that the validity of the time sharing scheme used in the proof of the direct part of \rPrp{iindayo2} (see \rSec{DistRand}) follows from the fact that the encoding unitaries are in the form of the tensor product as \req{randunitTP}.

\subsection{Comparison with Previous Results}

Distributed encoding can be viewed as a Shannon theoretical generalization of local encoding \cite{tanaka2007local}, which is a task of encoding classical information to a multipartite pure quantum state by local unitary operations.
Ref.~\cite{tanaka2007local} considered a one-shot and deterministic setting, and raised a question of whether it is possible to encode the maximum amount of information, i.e., the one determined by the dimension of the entire system, only by local unitary operations.
Ref.~\cite{tanaka2007local} proved that this is possible for several classes of pure states, while it has been left open whether this is possible for {\it all} multipartite pure states.
\rPrp{iindayo1} above answers this question in the affirmative in the setting of asymptotic limit of infinitely many copies and vanishingly small error.
Note that when $B$ and $E$ are trivial (one-dimensional) systems and $\rho$ is a pure state, then $ \hat{C}([Z])_\rho=\sum_{z\in[Z]}\log{d_{A_z}}$.

Refs.~\cite{bruss2004distributed,bruss2006dense} introduced a task called {\it distributed quantum dense coding}, which is a generalization of quantum superdense coding to a multi-sender scenario.
They considered both of the case with one receiver and the case with multiple receivers.
In the former, distributed quantum dense coding is similar to distributed encoding, i.e., the senders encode classical information to a multipartite quantum state only by local unitary operations. 
They proved that the maximum value of the Holevo information between the quantum system and the classical information to be encoded is equal to $C_{[Z]}^*$ defined by \req{dfnChat}.
\rPrp{iindayo1} above and its proof in \rSec{DistEnc} improves upon this result in that (i) \rPrp{iindayo1} not only deals with the total amount of information but also clarifies the trade-off relation between the maximum amounts of information that can be encoded by each of the senders, and (ii) in the proof in \rSec{DistEnc}, we explicitly show the existence of encoding operations and a decoding measurement that achieve the optimal encoding rate.
In the case of only one sender, \rPrp{iindayo1} reduces to the result of \cite{hiroshima2001optimal}.

Distributed randomization is similar to but is different from multi-sender decoupling \cite{chakraborty2021one} (see also Section 10 in \cite{ADHW2009}). 
In multi-sender decoupling, the goal is simply to destroy the correlation between $A_1\cdots A_{[Z]}$ and $W$.
In distributed randomization, we need not only to destroy the correlation between $A_1\cdots A_{[Z]}$ and $W$ but also to destroy the correlation among $A_1,\cdots ,A_{[Z]}$ and completely randomize these subsystems.
Thus, distributed randomization costs more randomness than multi-sender decoupling in general.
We remark that Ref.~\cite{dutil2011multiparty} considered distributed randomization that uses random projective measurements instead of random unitary operations (see Section 3.2.3 therein).

\subsection{Properties of The Set Functions $\hat{C}$ and $\hat{D}$}
\lsec{submodularity}

We describe properties of the set functions $\hat{C}$ and $\hat{D}$, defined by \req{dfnChat} and \req{dfnDhat}, respectively.
The properties will be used to prove the main results in \rSec{prfmain}, \rsec{DistEnc} and \rsec{DistRand}.
We define
\alg{
\check{D}(\Gamma)_\rho
&:=2\sum_{z\in\Gamma}\log{d_{A_z}}-\hat{D}(\Gamma)_\rho
\laeq{dfnDchec}
\\
&=\sum_{z\in\Gamma}\log{d_{A_z}}+S(A_\Gamma|W)_\rho.
\laeq{dfnDcheck}
}

\blmm{submodCDf}
The set functions $ \hat{C}(\Gamma)_\rho$ and $ \check{D}(\Gamma)_\rho$ are zero for the empty set, nonnegative, nondecreasing and strongly subadditive.
I.e., for $f=\hat{C},\check{D}$ and for any subsets $\Gamma,\Gamma',\Gamma''\subseteq[Z]$ satisfying $\Gamma\subseteq\Gamma'\subseteq[Z]$, it holds that
\alg{
f(\emptyset)_\rho&=0,
\laeq{zerof}
 \\
f(\Gamma)_\rho&\geq0,
\laeq{nnf}
 \\
f(\Gamma)_\rho&\leq f({\Gamma'})_\rho,\laeq{nondecCDf}
  \\
f(\Gamma)_\rho+f(\Gamma'')_\rho&\geq f(\Gamma\cup\Gamma'')_\rho+ f(\Gamma\cap\Gamma'')_\rho.\laeq{submodCDf}
}
\elmm

\bprf
Equality \req{zerof} immediately follows from the definitions of $\hat{C}$ and $\check{D}$.
Inequality \req{nnf} follows from the fact that the conditional entropy of a state $\varrho\in\ca{S}(\ca{H}^{AR})$ is bounded by the system dimension as $-\log{d_A}\leq S(A|R)_\varrho\leq\log{d_A}$.
To prove \req{nondecCDf} and \req{submodCDf}, note that $A_{\tilde\Gamma} A_{{\tilde\Gamma}_c}=A_{[Z]}$ for any $\tilde{\Gamma}\subseteq[Z]$.
From the definition of the conditional entropy, we have
\alg{
&
S(A_{\Gamma'}|A_{\Gamma_c'}V)_\rho
-
S(A_{\Gamma}|A_{\Gamma_c}V)_\rho
\nn\\
&=
S(A_{\Gamma_c}V)_\rho
-
S(A_{\Gamma_c'}V)_\rho
\\
&
=S(A_{\Gamma_c\backslash\Gamma_c'}|A_{\Gamma_c'}V)_\rho\\
&
=S(A_{\Gamma'\backslash\Gamma}|A_{\Gamma_c'}V)_\rho\\
&
\leq
\sum_{z\in\Gamma'\backslash\Gamma}\log{d_{A_z}}\\
&
=
\sum_{z\in\Gamma'}\log{d_{A_z}}
-
\sum_{z\in\Gamma}\log{d_{A_z}},
}
which implies \req{nondecCDf} for $\hat{C}$.
In the same way, we have
\alg{
&
S(A_{\Gamma'}|W)_\rho
-
S(A_{\Gamma}|W)_\rho
\nn\\
&=
S(A_{\Gamma'}W)_\rho
-
S(A_{\Gamma}W)_\rho
\\
&
=S(A_{\Gamma'\backslash\Gamma}|A_{\Gamma}W)_\rho\\
&
\geq
-\sum_{z\in\Gamma'\backslash\Gamma}\log{d_{A_z}}\\
&
=
\sum_{z\in\Gamma}\log{d_{A_z}}
-
\sum_{z\in\Gamma'}\log{d_{A_z}},
}
which implies \req{nondecCDf} for $\check{D}$.
We also have
\alg{
&\left( \hat{C}(\Gamma)_\rho+\hat{C}(\Gamma'')_\rho\right)
-
\left(\hat{C}(\Gamma\cup\Gamma'')_\rho+ \hat{C}(\Gamma\cap\Gamma'')_\rho\right)
\nn\\
&=
S(A_{\Gamma\cup\Gamma''}|A_{(\Gamma\cup\Gamma'')_c}V)_\rho
+
S(A_{\Gamma\cap\Gamma''}|A_{(\Gamma\cap\Gamma'')_c}V)_\rho
\nn\\
&\quad\quad
-S(A_\Gamma|A_{\Gamma_c}V)_\rho-S(A_{\Gamma''}|A_{\Gamma_c''}V)_\rho
\\
&=
S(A_{\Gamma_c}V)_\rho+S(A_{\Gamma_c''}V)_\rho
\nn\\
&\quad\quad
-
S(A_{\Gamma_c\cap\Gamma_c''}V)_\rho
-
S(A_{\Gamma_c\cup\Gamma_c''}V)_\rho.
}
The last line is nonnegative due to the strong subadditivity of the von Neumann entropy, which implies \req{submodCDf} for $\hat{C}$.
Similarly, we have
\alg{
&
S(A_\Gamma|W)_\rho
+
S(A_{\Gamma''}|W)_\rho
\nn\\
&
\quad
=
S(A_\Gamma W)_\rho
+
S(A_{\Gamma''}W)_\rho
-2S(W)_\rho
\\
&\quad
\geq
S(A_{\Gamma\cup\Gamma''}W)_\rho
+
S(A_{\Gamma\cap\Gamma''}W)_\rho
-2S(W)_\rho
\\
&\quad
=
S(A_{\Gamma\cup\Gamma''}|W)_\rho
+
S(A_{\Gamma\cap\Gamma''}|W)_\rho,
}
which implies \req{submodCDf} for $\check{D}$.
\QED
\eprf

\noindent
The following corollary immediately follows from \rLmm{submodCDf} and \req{dfnDchec}:

\bcrl{supmodDD}
The set function $\hat{D}(\Gamma)_\rho$ is zero for the empty set and strongly superadditive, i.e., $\hat{D}(\emptyset)=0$ and for any subsets $\Gamma,\Gamma'\subseteq[Z]$, it holds that 
\alg{
\hat{D}(\Gamma)_\rho+\hat{D}(\Gamma')_\rho&\leq \hat{D}(\Gamma\cup\Gamma')_\rho+ \hat{D}(\Gamma\cap\Gamma')_\rho.\laeq{submodD}
}
\ecrl

\noindent
Due to \rLmm{submodCDf}, the extremal points of the regions defined by \req{dontasp} and \req{dontask} are identified as follows:

\blmm{polymatvertex}
The region defined by \req{dontasp} is a polymatroid in $\bb{R}^Z$ and every extremal point $(C_1^*,\cdots,C_{[Z]}^*)$ is represented as
\alg{
\!
C_{\sigma(z)}^*
\!=\!
\log{d_{A_{\sigma(z)}}}
\!\!-\!
S(A_{\sigma(z)}|A_{\sigma(1)}\!\cdots\! A_{\sigma(z-1)}V)_{\rho},\!
\laeq{jasmine1}
}  
where $\sigma$ is a permutation on $[Z]$.
Similarly, the region defined by \req{dontask} is a contrapolymatroid in $\bb{R}^Z$ and every extremal point $(D_1^*,\cdots,D_{[Z]}^*)$ is represented as
\alg{
\!
D_{\sigma(z)}^*
\!=\!
\log{d_{A_{\sigma(z)}}}
\!\!-\!
S(A_{\sigma(z)}|A_{\sigma(1)}\!\cdots\! A_{\sigma(z-1)}W)_{\rho}.\!
\laeq{dsigmaz}
}  
\elmm

\bprf
Due to \rLmm{submodCDf} and the property of polymatroids (see Section 18.4 in \cite{welsh2010matroid}),
the extremal points of the region defined by \req{dontasp} are all points $(C_1^*,\cdots,C_{[Z]}^*)$ that can be represented as
\alg{
\!\!
C_{\sigma(z)}^*
&=
\hat{C}(\{\sigma(z')\}_{z'\in[z]})-\hat{C}(\{\sigma(z')\}_{z'\in[z-1]})
\nn\\
&=
\log{d_{A_{\sigma(z)}}}
\!\!-\!
S(A_{\sigma(z)}|A_{\sigma(z+1)}\!\cdots\! A_{\sigma(Z)}W)_{\rho},\!
}
where $\sigma$ is a permutation on $[Z]$.
Combining this with permutation of $[Z]$ in the inverse order, we obtain \req{jasmine1}. 
In the same way, the set of all tuples $(\tilde{D}_1,\cdots,\tilde{D}_Z)\in\mbb{R}^Z$ that satisfy the condition
$
\sum_{z\in\Gamma}\tilde{D}_z
\leq
\check{D}(\Gamma)_\rho
$
for any $\Gamma\in[Z]$ is a polymatroid, and all of its vertices $(\tilde{D}_1^*,\cdots,\tilde{D}_{[Z]}^*)$ are given by
\alg{
\!\!\!
\tilde{D}_{\sigma(z)}^*
&=
\check{D}(\{\sigma(z')\}_{z'\in[z]})
-\check{D}(\{\sigma(z')\}_{z'\in[z-1]})
\\
&
=
\log{d_{A_{\sigma(z)}}}
\!\!+\!
S(A_{\sigma(z)}|A_{\sigma(1)}\!\cdots\! A_{\sigma(z-1)}W)_{\rho}\!
}  
for some permutation $\sigma$.
Noting that $\check{D}$ is defined by \req{dfnDchec}, the vertices of the region defined by \req{dontask} is represented as $D_{\sigma(z)}^*=2\log{d_{A_{\sigma(z)}}}-\tilde{D}_{\sigma(z)}^*$ and are equal to \req{dsigmaz}.
\QED
\eprf

\noindent
We remark that it is not possible to identify the extremal points of the region defined by \req{maincond} along the same argument as in \rLmm{polymatvertex}.
This is because the function $I(A_\Gamma:A_{\Gamma_c}B|E)_\rho$ is not a nondecreasing function of $\Gamma$, i.e., it does not satisfy the condition \req{nondecCDf} in \rLmm{submodCDf}.

\section{Proof of The Main Theorem}
\lsec{prfmain}

In this section, we provide a proof of the main theorem (\rThm{mainthm} in \rSec{formandres}).
The proof of the converse part is based on a standard calculation of the entropies and the mutual informations.
For the direct part, we invoke the direct part of the distributed encoding theorem and the distributed randomization theorem (\rPrp{iindayo1} and \rPrp{iindayo2}).

\subsection{Proof of The Converse Part}
Suppose that a rate triplet $(R_1,\cdots,R_Z)$ is achievable for the state $\rho\in\ca{S}(\ca{H}^{A_1\cdots A_ZBE})$.
Fix arbitrary $\epsilon,\vartheta>0$, choose sufficiently large $n$ and let $M_z=2^{nR_z}$ for each $z\in[Z]$.
By definition, there exist a quantum system $A_z'$, a set of encoding CPTP maps $\mfk{E}_z\equiv\{\ca{E}_{z,m_z}\}_{m_z=1}^{M_z}$ from $A_z^n$ to $A_z'$ for each $z$ and a measurement $\{\Lambda_{{\bf m}}\}_{{\bf m}\in \bf{M}}$ on $A_1'\cdots A_Z'B^nE^n$ such that
\alg{
1-\frac{1}{|{\bf M}|}
\sum_{{\bf m}\in {\bf M}}
{\rm Tr}[\rho_{{\bf m}}\Lambda_{{\bf m}}]
\leq
\epsilon
\laeq{convreliable}
}
and
\alg{
\frac{1}{|{\bf M}|}
\sum_{{\bf m}\in {\bf M}}
\left\|\rho_{{\bf m}}^{A_1'\cdots A_Z'E^n}-\bar{\rho}^{A_1'\cdots A_Z'E^n}\right\|_1
\leq
\vartheta,
\laeq{convsecret}
}
where
\alg{
\rho_{{\bf m}}
:=
\left(\bigotimes_{z=1}^Z\ca{E}_{z,m_z}^{A_z^n\rightarrow A_z'}\right)
(\rho^{\otimes n})
\laeq{rhovecm0II}
}
and
\alg{
\bar{\rho}
:=
\frac{1}{|{\bf M}|}
\sum_{{\bf m}\in {\bf M}}
\rho_{{\bf m}}.
\laeq{rhobar0II}
}

Fix an arbitrary subset $\Gamma\subseteq[Z]$.
Let $A_\Gamma'$ and $A_{\Gamma_c}'$ denote the systems composed of $\{A_z'\}_{z\in\Gamma}$ and $\{A_z'\}_{z\in\Gamma_c}$, respectively.
The set ${\bf M}$ is divided into ${\bf M}_\Gamma\times{\bf M}_{\Gamma_c}$, where
\alg{
{\bf M}_\Gamma:=\bigtimes_{z\in\Gamma}[M_z],\;
{\bf M}_{\Gamma_c}:=\bigtimes_{z\in\Gamma_c}[M_z].
}
Correspondingly, each element ${\bf m}\in{\bf M}$ is represented as ${\bf m}=({\bf m}_{\Gamma},{\bf m}_{\Gamma_c})$ by ${\bf m}_{\Gamma}\in{\bf M}_\Gamma$ and ${\bf m}_{\Gamma_c}\in{\bf M}_{\Gamma_c}$.
Let $M_\Gamma$ and $M_{\Gamma_c}$ be quantum systems with fixed orthonormal bases $\{|{\bf m}_\Gamma\rangle\}_{{\bf m}_\Gamma\in{\bf M}_\Gamma}$ and $\{\ket{{\bf m}_{\Gamma_c}}\}_{{\Gamma_c}\in{\bf M}_{\Gamma_c}}$, respectively. 
We define CPTP maps $\ca{E}_{\Gamma,{\bf m}_\Gamma}:A_\Gamma^n\rightarrow A_\Gamma'$ and $\ca{E}_{\Gamma_c,{\bf m}_{\Gamma_c}}:A_{\Gamma_c}^n\rightarrow A_{\Gamma_c}'$ by
\alg{
\ca{E}_{\Gamma,{\bf m}_\Gamma}:=\bigotimes_{z\in\Gamma}\ca{E}_{z,m_z},
\;
\ca{E}_{\Gamma_c,{\bf m}_{\Gamma_c}}:=\bigotimes_{z\in\Gamma_c}\ca{E}_{z,m_z}.
}
We also define $\tilde{\ca{E}}_\Gamma:A_\Gamma^n\rightarrow A_\Gamma'M_\Gamma$ and $\tilde{\ca{E}}_{\Gamma_c}:A_{\Gamma_c}^n\rightarrow A_{\Gamma_c}'M_{\Gamma_c}$ by 
\alg{
\tilde{\ca{E}}_\Gamma(\cdot)
&
:=
\frac{1}{|{\bf M}_\Gamma|}\sum_{{\bf m}_\Gamma\in{\bf M}_\Gamma}
\proj{{\bf m}_\Gamma}^{M_\Gamma}\otm
\ca{E}_{\Gamma,{\bf m}_\Gamma}(\cdot),
\nn\\
\tilde{\ca{E}}_{\Gamma_c}(\cdot)
&
:=
\frac{1}{|{\bf M}_{\Gamma_c}|}\sum_{{\bf m}_{\Gamma_c}\in{\bf M}_{\Gamma_c}}
\proj{{\bf m}_{\Gamma_c}}^{M_{\Gamma_c}}\otm
\ca{E}_{\Gamma_c,{\bf m}_{\Gamma_c}}(\cdot).
\nn
}
The state after the encoding operation is represented by
\alg{
\tilde{\rho}
=
\tilde{\ca{E}}_{\Gamma}\otm\tilde{\ca{E}}_{\Gamma_c}(\rho^{\otm n}).
\laeq{elelm}
}
It is straightforward to verify that 
\alg{
\bar{\rho}={\rm Tr}_{M_\Gamma M_{\Gamma_c}}[\tilde{\rho}],
\quad
\pi^{M_{\Gamma}M_{\Gamma_c}}={\rm Tr}_{A_{[Z]}'B^nE^n}[\tilde{\rho}].
\laeq{jiji2}
}
Noting that both $\tilde{\rho}$ and $\pi^{M_{\Gamma}M_{\Gamma_c}}$ are diagonal in $\{|{\bf m}_\Gamma\rangle\}$ and $\{\ket{{\bf m}_{\Gamma_c}}\}$,
the secrecy condition \req{convsecret} is equivalent to
\alg{
\left\|\tilde{\rho}^{M_\Gamma M_{\Gamma_c}A_{\Gamma}'A_{\Gamma_c}'E^n}\!-\!\pi^{M_\Gamma M_{\Gamma_c}}\!\otm\!\bar{\rho}^{A_{\Gamma}'A_{\Gamma_c}'E^n}\right\|_1
\leq
\vartheta.
\!
}
Tracing out $M_{\Gamma_c}A_{\Gamma_c}'$, we obtain
\alg{
\left\|\tilde{\rho}^{M_\Gamma A_{\Gamma}'E^n}\!-\!\pi^{M_\Gamma}\!\otm\!\bar{\rho}^{A_{\Gamma}'E^n}\right\|_1
\leq
\vartheta.
\!
\laeq{convsecret2}
}

We calculate the entropies and the mutual informations of $\tilde{\rho}$ along the same line of Appendix C in \cite{sharma2020conditional}.
First, let $M_\Gamma'$ be the system to which the $\Gamma$ part of the decoding result is registered.
Due to the reliability condition \req{convreliable} and Fano's inequality (see e.g.~Theorem 2.10.1 in \cite{cover05}), we have
\alg{
I(M_\Gamma:M_\Gamma')
&
=
S(M_\Gamma)
-
S(M_\Gamma|M_\Gamma')
\nn\\
&
\geq
(1-\epsilon)\log{|{\bf M}_\Gamma|}-h(\epsilon)
\\
&=
n(1-\epsilon)\sum_{z\in\Gamma}R_z-h(\epsilon).
\laeq{yatto}
}
Second, from \req{jiji2}, \req{convsecret2} and the Alicki-Fannes inequality (\!\!\cite{alicki04}: see \cite{winter2016tight} for an improved version), we obtain
\alg{
&
I(M_\Gamma :A_\Gamma'E^n)_{\tilde{\rho}}
\nn\\
&=
S(M_\Gamma )_{\pi}
-
S(M_\Gamma |A_\Gamma'E^n)_{\tilde{\rho}}
\\
&=
S(M_\Gamma |A_\Gamma'E^n)_{\pi\otm\bar{\rho}}
-
S(M_\Gamma |A_\Gamma'E^n)_{\tilde{\rho}}
\\
&
\leq
2\vartheta\log{d_{M_\Gamma}}+(1+\vartheta)h\left(\frac{\vartheta}{1+\vartheta}\right)
\\
&
=
2n\vartheta\sum_{z\in\Gamma}R_z+(1+\vartheta)h\left(\frac{\vartheta}{1+\vartheta}\right).
\laeq{yatto2}
}
Third, we have
\alg{
&
I(M_\Gamma:M_\Gamma')
-
I(M_\Gamma :A_\Gamma'E^n)_{\tilde{\rho}}
\laeq{yuki-1}\\
&\leq
I(M_\Gamma:A_\Gamma' A_{\Gamma_c}'B^nE^n)_{\tilde{\rho}}
-
I(M_\Gamma:A_\Gamma'E^n)_{\tilde{\rho}}
\laeq{yuki0}\\
&=
I(M_\Gamma:A_{\Gamma_c}'B^n|A_\Gamma' E^n)_{\tilde{\rho}}
\laeq{yuki1}
\\
&\leq
I(M_\Gamma:A_{\Gamma_c}^nB^n|A_\Gamma' E^n)_{\tilde{\ca{E}}_{\Gamma}(\rho^{\otm n})}
\laeq{yuki2}
\\
&=
\sum_{i=1}^nI(M_\Gamma:A_{\Gamma_c,i}B_i|A_\Gamma' A_{\Gamma_c}^{i-1}B^{i-1}E^n)_{\tilde{\ca{E}}_{\Gamma}(\rho^{\otm n})}
\laeq{yuki3}
\\
&=
\sum_{i=1}^n\left[I(M_\Gamma A_\Gamma' A_{\Gamma_c}^{i-1}\!B^{i-1}\!E^n_{\backslash i}\!:\!A_{\Gamma_c,i}B_i|E_i)_{\tilde{\ca{E}}_{\Gamma}(\rho^{\otm n})}\right.\nn\\
&\quad\quad\left.-I(A_\Gamma' A_{\Gamma_c}^{i-1}\!B^{i-1}\!E^n_{\backslash i}\!:\!A_{\Gamma_c,i}B_i|E_i)_{\tilde{\ca{E}}_{\Gamma}(\rho^{\otm n})}\right]
\laeq{yuki4}
\\
&\leq
\sum_{i=1}^nI(M_\Gamma A_\Gamma' A_{\Gamma_c}^{i-1}\!B^{i-1}\!E^n_{\backslash i}\!:\!A_{\Gamma_c,i}B_i|E_i)_{\tilde{\ca{E}}_{\Gamma}(\rho^{\otm n})}
\laeq{yuki5}
\\
&\leq
\sum_{i=1}^nI(A_\Gamma^n A_{\Gamma_c}^{i-1}B^{i-1}E^n_{\backslash i}:A_{\Gamma_c,i}B_i|E_i)_{\rho^{\otm n}}
\laeq{yuki6}
\\
&=
\sum_{i=1}^nI(A_{\Gamma,i}:A_{\Gamma_c,i}B_i|E_i)_{\rho^{\otm n}}
\laeq{yuki7}
\\
&=
nI(A_\Gamma:A_{\Gamma_c}B|E)_{\rho}.
\laeq{yatto3}
}
Here, 
\req{yuki0} follows from the data processing inequality of the mutual information and the fact that $M_\Gamma'$ is obtained by performing a measurement on $A_\Gamma' A_{\Gamma_c}'B^nE^n$;
\req{yuki1} due to the chain rule of the mutual information; 
\req{yuki2} from \req{elelm} and the data processing inequality; 
\req{yuki3} from the chain rule of the mutual information, where $A_{\Gamma_c}^{i-1}$ and $B^{i-1}$ denotes the systems $A_{\Gamma_c,1}\cdots A_{\Gamma_c,i-1}$ and $B_1\cdots B_{i-1}$, respectively;
\req{yuki4} due to the chain rule of the mutual information, where $E^n_{\backslash i}$ denotes $E_1\cdots E_{i-1}E_{i+1}\cdots E_n$; 
\req{yuki5} from the non-negativity of the conditional mutual information; 
\req{yuki6} from the data processing inequality; 
\req{yuki7} because $\rho^{\otm n}$ is a product state between $A_{\Gamma,i}A_{\Gamma_c,i}B_iE_i$ and the other systems; and \req{yatto3} because the state on $A_{\Gamma,i}A_{\Gamma_c,i}B_iE_i$ is $\rho$ for each $i$.
Substituting \req{yatto} and \req{yatto2} into \req{yuki-1},
we obtain
\alg{
&
n(1-\epsilon-2\vartheta)\sum_{z\in\Gamma}R_z
\nn\\
&\;
\leq
nI(A_\Gamma\!:\!A_{\Gamma_c}B|E)_{\rho}
\!+\!
h(\epsilon)
\!+\!
(1\!+\!\vartheta)h\left(\frac{\vartheta}{1\!+\!\vartheta}\right)\!.\!
}
Since this relation holds for any $\epsilon,\vartheta>0$ and sufficiently large $n$, we arrive at
\alg{
\sum_{z\in\Gamma}R_z
\leq
I(A_\Gamma:A_{\Gamma_c}B|E)_{\rho}.
}
Noting that the above inequality holds for any $\Gamma\subseteq[Z]$, we complete the proof of the converse part.
\QED

\subsection{Proof of The Direct Part}

We apply the distributed encoding theorem (\rPrp{iindayo1}) and the distributed randomization theorem (\rPrp{iindayo2}) under the correspondence $V\rightarrow BE$ and $W\rightarrow E$.
Recall that $\hat{C}(\Gamma)_\rho$ and $\hat{D}(\Gamma)_\rho$ are defined by \req{dfnChat} and \req{dfnDhat}, respectively, which yields
\alg{
\hat{C}(\Gamma)_\rho&=\sum_{z\in\Gamma}\log{d_{A_z}}-S(A_\Gamma|A_{\Gamma_c}BE)_\rho,
\\
\hat{D}(\Gamma)_\rho&=\sum_{z\in\Gamma}\log{d_{A_z}}-S(A_\Gamma|E)_\rho.
}
It is straightforward to verify that
\alg{
 \hat{C}(\Gamma)_\rho-\hat{D}(\Gamma)_\rho
=
I(A_\Gamma:A_{\Gamma_c}B|E)_\rho.
}
Thus, the condition \req{maincond} is equivalent to
\alg{
\sum_{z\in\Gamma}R_z\leq \hat{C}(\Gamma)_\rho-\hat{D}(\Gamma)_\rho.
\laeq{aiwo}
}
We prove the direct part of \rThm{mainthm} based on the following lemma:

\blmm{tupleRCD}
For any rate tuple $(R_1,\cdots,R_Z)$ that satisfies the condition
\alg{
\sum_{z\in\Gamma}R_z< \hat{C}(\Gamma)_\rho-\hat{D}(\Gamma)_\rho,
\laeq{aiwowo}
}
 there exists a pair of rate tuples $(C_1,\cdots,C_Z)$ and $(D_1,\cdots,D_Z)$ that satisfy
 \alg{
 \sum_{z\in\Gamma}C_z< \hat{C}(\Gamma)_\rho,
 \quad
 \sum_{z\in\Gamma}D_z>\hat{D}(\Gamma)_\rho,
 \laeq{inin1}
 }
 respectively, for any $\Gamma\in[Z]$, and it holds that 
 \alg{
 R_z=C_z-D_z
  \laeq{inin2}
 }
 for all $z\in[Z]$.
\elmm

\noindent
A proof of \rLmm{tupleRCD} will be provided at the end of this subsection.

To prove the direct part of \rThm{mainthm},
let $(R_1,\cdots,R_Z)$ be an arbitrary rate tuple that satisfies the condition \req{aiwowo}.
Fix a pair of rate tuples $(C_1,\cdots,C_Z)$ and $(D_1,\cdots,D_Z)$ that satisfy \req{inin1} and \req{inin2}.
Fix arbitrary $\epsilon,\vartheta,\xi>0$ and choose sufficiently large $n$.
Let $K_z\equiv 2^{nC_z}$, $L_z\equiv 2^{nD_z}$ and $M_z\equiv 2^{nR_z}$ for each $z$.
Note that $K_z=L_zM_z$.
Let $U_{z,k_z,i}\in\ca{U}(\ca{H}^{A_z})$ for $i\in[n]$ and $U_{z,k_z}:=\bigotimes_{i=1}^nU_{z,k_z,i}$.
We let $A_z'=A_z^n$ and construct the encoding operation $\{\ca{E}_{z,m_z}\}_{m_z=1}^{M_z}$ by
\alg{
\ca{E}_{z,m_z}(\cdot)
=
\frac{1}{L_z}\sum_{k_z=(m_z-1)L_z+1}^{m_zL_z}U_{z,k_z}(\cdot)U_{z,k_z}^\dagger.
\laeq{rerere}
}
In the following, we prove that if we choose $U_{z,k_z,i}$ randomly and independently according to the Haar measure for each $i\in[n]$, $k_z\in[K_z]$ and $z\in[Z]$, the set of encoding operations constructed as \req{rerere} satisfies both the reliability condition and the secrecy condition with a probability greater than $1-2\xi$.
Recall that the state after the encoding operation corresponding to the message value ${\bf m}$ is given by
\alg{
\rho_{{\bf m}}
:=
\left(\bigotimes_{z=1}^Z\ca{E}_{z,m_z}^{A_z^n}\right)
(\rho^{\otimes n}).
\laeq{rerer2}
}

To prove the reliability condition, define
$
U_{{\bf k}}
=
\bigotimes_{z=1}^ZU_{z,k_z}
$
and $\rho_{{\bf k}}:=(U_{{\bf k}}\otm I^{B^nE^n})\rho^{\otimes n}(U_{{\bf k}}\otm I^{B^nE^n})^\dagger$ for ${\bf k}\equiv(k_1,\cdots,k_Z)$.
Due to the distributed encoding theorem (\rPrp{iindayo1}), with probability no smaller than $1-\xi$,
there exists a measurement $\{\Lambda_{{\bf k}}\}_{{\bf k}\in{\bf K}}$ on $A_1^n\cdots A_Z^nB^nE^n$ and it holds that
\alg{
1-\frac{1}{|{\bf K}|}
\sum_{{\bf k}\in {\bf K}}
{\rm Tr}[\rho_{{\bf k}}\Lambda_{{\bf k}}]
\leq
\epsilon.
}
We construct the decoding measurement by 
\alg{
\Lambda_{\bf m}
:=
\sum_{k_1=(m_1-1)L_1+1}^{m_1L_1}\cdots\sum_{k_Z=(m_Z-1)L_Z+1}^{m_ZL_Z}\Lambda_{{\bf k}}.
}
It is straightforward to verify that
\alg{
\frac{1}{|{\bf M}|}
\sum_{{\bf m}\in {\bf M}}
\Lambda_{{\bf m}}
=
\frac{1}{|{\bf K}|}
\sum_{{\bf k}\in {\bf K}}
\Lambda_{{\bf k}}
=I,
}
thus $\{\Lambda_{\bf m}\}_{{\bf m}\in {\bf M}}$ is indeed a POVM.
Noting that $\rho_{{\bf m}}$ in \req{rerer2} is calculated to be
\alg{
\rho_{{\bf m}}
=
\frac{1}{L_1\cdots L_Z}\sum_{k_1=(m_1-1)L_1+1}^{m_1L_1}\cdots\sum_{k_Z=(m_Z-1)L_Z+1}^{m_ZL_Z}\rho_{{\bf k}},
}
we have
\alg{
&
{\rm Tr}[\rho_{{\bf m}}\Lambda_{{\bf m}}]
\nn\\
&\;
\geq
\frac{1}{L_1\cdots L_Z}\sum_{k_1=(m_1-1)L_1+1}^{m_1L_1}\!\!\cdots\!\!\sum_{k_Z=(m_Z-1)L_Z+1}^{m_ZL_Z}\!{\rm Tr}[\rho_{{\bf k}}\Lambda_{{\bf k}}].\!
}
Thus
\alg{
\frac{1}{|{\bf M}|}
\!\sum_{{\bf m}\in {\bf M}}
\!{\rm Tr}[\rho_{{\bf m}}\Lambda_{{\bf m}}]
\geq
\frac{1}{|{\bf K}|}
\!\sum_{{\bf k}\in {\bf K}}
\!{\rm Tr}[\rho_{{\bf k}}\Lambda_{{\bf k}}]
\geq
1\!-\!\epsilon,\!
}
which implies the reliability condition.

To prove the secrecy condition, we evaluate the information leakage as
\alg{
&
\frac{1}{|{\bf M}|}
\sum_{{\bf m}\in {\bf M}}
\left\|\rho_{{\bf m}}^{A_{[Z]}^nE^n}-\bar{\rho}^{A_{[Z]}^nE^n}\right\|_1
\nn
\\
&\leq
\frac{1}{|{\bf M}|}
\sum_{{\bf m}\in {\bf M}}
\left\|\rho_{{\bf m}}^{A_{[Z]}^nE^n}-\pi^{A_{[Z]}^n}\otm\bar{\rho}^{E^n}\right\|_1
\nn\\
&\quad\quad+\left\|\bar{\rho}^{A_{[Z]}^nE^n}-\pi^{A_{[Z]}^n}\otm\bar{\rho}^{E^n}\right\|_1
\laeq{67}
\\
&\leq
\frac{2}{|{\bf M}|}
\sum_{{\bf m}\in {\bf M}}
\left\|\rho_{{\bf m}}^{A_{[Z]}^nE^n}-\pi^{A_{[Z]}^n}\otm\bar{\rho}^{E^n}\right\|_1,
\laeq{68}
}
where \req{67} follows from the triangle inequality and \req{68} from the convexity of the trace distance.
Hence, by the distributed randomization theorem (\rPrp{iindayo2}), we have
\alg{
\frac{1}{|{\bf M}|}
\sum_{{\bf m}\in {\bf M}}
\left\|\rho_{{\bf m}}^{A_{[Z]}^nE^n}-\bar{\rho}^{A_{[Z]}^nE^n}\right\|_1
\leq
2\vartheta
}
with a probability no smaller than $1-\xi$.

In total, with a probability no smaller than $(1-\xi)^2>1-2\xi$, the reliability condition and the secrecy condition are both satisfied.
Since this relation holds for any $\epsilon,\vartheta,\xi>0$ and sufficiently large $n$, we conclude that any rate tuple satisfying the condition \req{aiwowo} is achievable.
Taking the closure of the region defined by \req{aiwowo}, we obtain \req{aiwo} and complete the proof of the direct part.
\QED

It remains to prove \rLmm{tupleRCD}.
Let $S$ be a finite set.
A function $f:2^S\rightarrow\mbb{R}$ is said to be a {\it submodular function} if it holds that
\alg{
f(A)+f(B)\geq f(A\cup B)+f(A\cap B)
}
for any $A,B\subseteq S$.
A function $g:2^S\rightarrow\mbb{R}$ is said to be a {\it supermodular function} if $-g$ is submodular.
We prove \rLmm{tupleRCD} based on the following lemma:

\blmm{separation}
Let $f:2^S\rightarrow\mbb{R}$ a submodular function and $g:2^S\rightarrow\mbb{R}$ be a supermodular function such that 
\alg{
&f(\emptyset)=g(\emptyset)=0,\laeq{fg0}\\
&g(A)\leq f(A)\quad(\forall A\subseteq S, A\neq\emptyset).\laeq{AAAS}
}
There exists $\{R_s\}_{s\in S}$ such that 
\alg{
g(A)\leq\sum_{s\in A}R_s\leq f(A)
\laeq{AAASII}
}
for any $A\subseteq S$ satisfying $A\neq\emptyset$.
If the condition \req{AAAS} is strict inequalities, both inequalities in \req{AAASII} can be strict inequalities.
\elmm

\bprf
The former statement was proved in Section 4 of \cite{lovasz1983submodular}.
To prove the latter statement, suppose that $g(A)< f(A)$ for any $A\subseteq S$ such that $A\neq\emptyset$.
Let
\alg{
\Delta:=\min_{A\subseteq S, A\neq\emptyset}\frac{1}{2|A|}(f(A)-g(A))
}
and define $f',g':2^S\rightarrow\mbb{R}$ by
\alg{
f'(A)=f(A)-|A|\Delta,
\quad
g'(A)=g(A)+|A|\Delta.
}
Applying the former statement to $f'$ and $g'$, we complete the proof.
\QED
\eprf

\!\!\!{\bf Proof of \rLmm{tupleRCD}:}
Due to \rLmm{submodCDf} and \rCrl{supmodDD}, the set functions $ \hat{C}(\Gamma)_\rho-\sum_{z\in\Gamma}R_z$ and $\hat{D}(\Gamma)_\rho$ are submudular and supermodular, respectively, and are equal to zero for $\Gamma=\emptyset$.
From \req{aiwowo}, we have $\hat{D}(\Gamma)_\rho< \hat{C}(\Gamma)_\rho-\sum_{z\in\Gamma}R_z$.
Hence, due to \rLmm{separation}, there exists $\{D_z\}_{z\in[Z]}$ such that $\hat{D}(\Gamma)_\rho<\sum_{z\in\Gamma}D_z< \hat{C}(\Gamma)_\rho-\sum_{z\in\Gamma}R_z$ for any nonempty $\Gamma\subseteq[Z]$.
Letting $C_z=D_z+R_z$ completes the proof.
We remark that the same argument was used in \cite{chou2021private} to prove the achievability of the private classical capacity of a classical-quantum multiple-access channel (see Inequality (19) therein).
\QED

\section{Proof of \rPrp{iindayo1}}
\lsec{DistEnc}

In this section, we prove the distributed encoding theorem (\rPrp{iindayo1}).
We will use the same notations as in \rSec{DEThm}.

\subsection{Proof of The Converse Part}
Suppose that a triplet $(C_1,\cdots,C_Z)$ is achievable for distributed encoding on the state $\rho\in\ca{S}(\ca{H}^{A_1\cdots A_ZV})$.
Fix arbitrary $\epsilon>0$, choose sufficiently large $n$ and let $K_z=2^{nC_z}$.
By definition, there exist a set of unitaries $\mfk{U}_z\equiv\{U_{z,k_z}\}_{k_z=1}^{K_z}$ on $A_z^n$ for each $z$ and a measurement $\{\Lambda_{{\bf k}}\}_{{\bf k}\in {\bf K}}$ on $A_{[Z]}^nV^n$ such that
\alg{
1-
\frac{1}{|{\bf K}|}
\sum_{{\bf k}\in {\bf K}}
{\rm Tr}[\rho_{{\bf k}}\Lambda_{{\bf k}}]
\leq
\epsilon,
\laeq{measerrorII}
}
where
$
\rho_{{\bf k}}
:=
(U_{{\bf k}}\otm I^{V^n})\rho^{\otimes n}(U_{{\bf k}}\otm I^{V^n})^\dagger
$
and
$U_{{\bf k}}:=\bigotimes_{z=1}^ZU_{z,k_z}$.

Let $\Gamma\in[Z]$ be any subset.
The set ${\bf K}$ is represented as ${\bf K}_\Gamma\times{\bf K}_{\Gamma_c}$, where
${\bf K}_\Gamma:=\bigtimes_{z\in\Gamma}[K_z]$,
${\bf K}_{\Gamma_c}:=\bigtimes_{z\in\Gamma_c}[K_z]$.
Each element ${\bf k}\in{\bf K}$ is written as ${\bf k}=({\bf k}_{\Gamma},{\bf k}_{\Gamma_c})$ by ${\bf k}_{\Gamma}\in{\bf K}_\Gamma$ and ${\bf k}_{\Gamma_c}\in{\bf K}_{\Gamma_c}$.
We define $U_{{\bf k}_\Gamma}:=\bigotimes_{z\in\Gamma}U_{z,k_z}$ and $U_{{\bf k}_{\Gamma_c}}:=\bigotimes_{z\in\Gamma_c}U_{z,k_z}$, by which $U_{{\bf k}}$ is represented as $U_{{\bf k}_{\Gamma}}^{A_{\Gamma}}\otm U_{{\bf k}_{\Gamma_c}}^{A_{\Gamma_c}}$ and $\rho_{{\bf k}}$ as $\rho_{{\bf k}_{\Gamma},{\bf k}_{\Gamma_c}}$.
It is straightforward to verify that
\alg{
&{\rm Tr}_{A_\Gamma^n}[\rho_{{\bf k}_{\Gamma},{\bf k}_{\Gamma_c}}]
=
\rho_{{\bf k}_{\Gamma_c}}^{A_{\Gamma_c}^nV^n}
\nn\\
&\quad\quad
:=
(U_{{\bf k}_{\Gamma_c}}^{A_{\Gamma_c}}\!\otm\! I^{V^n})(\rho^{A_{\Gamma_c} V})^{\otm n}(U_{{\bf k}_{\Gamma_c}}^{A_{\Gamma_c}}\!\otm\! I^{V^n})^\dagger.
\laeq{erumoa}
}
Let $K_\Gamma$ and $K_{\Gamma_c}$ be the systems with fixed orthonormal bases $\{|{\bf k}_{\Gamma}\rangle\}_{{\bf k}_{\Gamma}}$ and $\{|{\bf k}_{\Gamma}\rangle\}_{{\bf k}_{\Gamma}}$, respectively.
The state after the encoding operation is represented by
\alg{
\!
\tilde{\rho}
\!=\!
\frac{1}{|{\bf K}|}\!\sum_{{\bf k}_{\Gamma},{\bf k}_{\Gamma_c}}\!\!\!\proj{{\bf k}_{\Gamma}}^{K_\Gamma}\!\otm\!\proj{{\bf k}_{\Gamma_c}}^{K_{\Gamma_c}}
\!\otm\!\rho_{{\bf k}_{\Gamma},{\bf k}_{\Gamma_c}}\!.\!\!
\laeq{naiki1}
}
Using \req{erumoa},
it is straightforward to verify that
\alg{
{\rm Tr}_{K_\Gamma A_\Gamma^n}[\tilde{\rho}]
\!=\!
\frac{1}{|{\bf K}_{\Gamma_c}|}\!\sum_{{\bf k}_{\Gamma_c}}\!\!\proj{{\bf k}_{\Gamma_c}}^{K_{\Gamma_c}}
\!\otm\!\rho_{{\bf k}_{\Gamma_c}}^{A_{\Gamma_c}^nV^n}\!.\!
\laeq{naiki2}
}

To prove the converse part, we calculate the conditional entropies of $\tilde{\rho}$.
We have
\alg{
&
S(K_\Gamma|A_{[Z]}^nV^n)_{\tilde{\rho}}
\laeq{SS0}\\
&\geq
S(K_\Gamma|A_{[Z]}^nV^nK_{\Gamma_c})_{\tilde{\rho}}
\laeq{SS1}
\\
&=
S(K_\Gamma K_{\Gamma_c})_{\tilde{\rho}}
+
S(A_{[Z]}^nV^n|K_\Gamma K_{\Gamma_c})_{\tilde{\rho}}
\nn\\
&\quad\quad
-
S(A_{[Z]}^nV^nK_{\Gamma_c})_{\tilde{\rho}}
\laeq{SS2}
\\
&\geq
S(K_\Gamma K_{\Gamma_c})_{\tilde{\rho}}
+
S(A_{[Z]}^nV^n|K_\Gamma K_{\Gamma_c})_{\tilde{\rho}}
+
S(K_{\Gamma_c})_{\tilde{\rho}}
\nn\\
&\quad\quad
-
S(A_{\Gamma_c}^nV^n K_{\Gamma_c})_{\tilde{\rho}}
-
S(A_\Gamma^n K_{\Gamma_c})_{\tilde{\rho}}
\laeq{SS3}\\
&=
S(K_\Gamma|K_{\Gamma_c})_{\tilde{\rho}}
+
S(A_{[Z]}^nV^n|K_{[Z]})_{\tilde{\rho}}
\nn\\
&\quad\quad
-
S(A_{\Gamma_c}^nV^n|K_{\Gamma_c})_{\tilde{\rho}}
-
S(A_\Gamma^n|K_{\Gamma_c})_{\tilde{\rho}},
\laeq{SS4}
}
where \req{SS1} follows from the data processing inequality; \req{SS2} from the chain rule; \req{SS3} from the strong subadditivity and $A_\Gamma A_{\Gamma_c}=A_{[Z]}$; and \req{SS4} from the definition of the conditional entropy and $K_\Gamma K_{\Gamma_c}=K_{[Z]}$.
Noting that $K_{\Gamma_c}$ is uncorrelated with $K_\Gamma$, the first term in \req{SS4} is evaluated as
\alg{
\!\!
S(K_\Gamma|K_{\Gamma_c})_{\tilde{\rho}}
=
S(K_\Gamma)_{\tilde{\rho}}
=
\sum_{z\in\Gamma}\log{K_z}
=
n\!\sum_{z\in\Gamma}\!C_z.\!\!
\laeq{SR1}
}
The forth term is bounded by the system dimension as
\alg{
&
S(A_\Gamma^n|K_{\Gamma_c})_{\tilde{\rho}}
\leq
\log{d_{A_\Gamma^n}}
=
n\sum_{z\in\Gamma}\log{d_{A_z}}.
\laeq{SR2}
}
Due to \req{naiki1} and the unitary invariance of the von Neumann entropy, the second term in \req{SS4} is evaluated to be
\alg{
&
S(A_{[Z]}^nV^n|K_{[Z]})_{\tilde{\rho}}
=
\frac{1}{|{\bf K}|}\!\sum_{{\bf k}\in{\bf K}}
S\left(\rho_{\bf k}^{A_{[Z]}^nV^n}\right)
\\
&
=
S\left((\rho^{A_{[Z]}V})^{\otm n}\right)
=
nS(A_{[Z]}V)_{\rho}.
\laeq{SR4}
}
Similarly, due to \req{naiki2} we have
\alg{
&
S(A_{\Gamma_c}^nV^n|K_{\Gamma_c})_{\tilde{\rho}}
=
\frac{1}{|{\bf K}_{\Gamma_c}|}\!\sum_{{\bf k}_{\Gamma_c}\in{\bf K}_{\Gamma_c}}
S\left(\rho_{{\bf k}_{\Gamma_c}}^{A_{\Gamma_c}^nV^n}\right)
\\
&
=
S\left((\rho^{A_{\Gamma_c}V})^{\otm n}\right)
=
nS(A_{\Gamma_c}V)_{\rho}.
\laeq{SR3}
}
Let $K_\Gamma'$ be the system to which the $\Gamma$ part of the decoding result is recorded.
Since $K_\Gamma'$ is obtained as a result of a measurement on $A_{[Z]}^nV^n$,
the data processing inequality yields
\alg{
S(K_\Gamma|K_\Gamma')
\geq
S(K_\Gamma|A_{[Z]}^nV^n)_{\tilde{\rho}}.
\laeq{SR5}
}
Furthermore, due to the condition \req{measerrorII} and Fano's inequality (see e.g.~Theorem 2.10.1 in \cite{cover05}), we have
\alg{
S(K_\Gamma|K_\Gamma')
&
\leq
\epsilon\log{|{\bf K}_\Gamma|}+h(\epsilon)
\nn\\
&=
n\epsilon\sum_{z\in\Gamma}C_z+h(\epsilon).
\laeq{SR6}
}
We substitute \req{SR1}, \req{SR2}, \req{SR4} and \req{SR3} all into \req{SS3}, and \req{SR5} and \req{SR6} into \req{SS0}.
Noting that $S(A_{[Z]}V)_{\rho}-S(A_{\Gamma_c}V)_{\rho}=S(A_\Gamma|A_{\Gamma_c}V)_{\rho}$, we obtain
\alg{
&
n(1-\epsilon)\sum_{z\in\Gamma}C_z
\nn\\
&\quad
\leq
n\left(\sum_{z\in\Gamma}\log{d_{A_z}}
-
S(A_\Gamma|A_{\Gamma_c}V)_{\rho}\right)+h(\epsilon).
}
Since this relation holds for any $\epsilon>0$ and any sufficiently large $n$, we arrive at
\alg{
\sum_{z\in\Gamma}C_z
\leq
\sum_{z\in\Gamma}\log{d_{A_z}}
-
S(A_\Gamma|A_{\Gamma_c}V)_{\rho}.
}
Noting that the above inequality holds for any $\Gamma\subseteq[Z]$,
we complete the proof of the converse part.
\QED

\subsection{Proof of The Direct Part}

Recall that the region defined by \req{dontasp} is a polymatroid in $\bb{R}^Z$ and every extremal point $(C_1^*,\cdots,C_{[Z]}^*)$ is represented as
\alg{
\!
C_{\sigma(z)}^*
\!=\!
\log{d_{A_{\sigma(z)}}}
\!\!-\!
S(A_{\sigma(z)}|A_{\sigma(1)}\!\cdots\! A_{\sigma(z-1)}V)_{\rho},\!
}  
where $\sigma$ is a permutation on $[Z]$ (see \rLmm{polymatvertex}).  
Due to the time-sharing scheme, it suffices to prove that all these extremal points are in the achievable rate region.
Without loss of generality, it suffices to prove that a rate tuple $(C_1,\cdots,C_{[Z]})$ is achievable if
\alg{
C_{z}
<
\log{d_{A_{z}}}
-
S(A_{z}|A_{1}\cdots A_{z-1}V)_{\rho}
\laeq{TTTe}
}  
for all $z\in[Z]$.
The proof is based on the following two lemmas:

\blmm{trans1}
Let $A$ and $Q$ be finite-dimensional quantum systems represented by Hilbert spaces $\ca{H}^A$ and $\ca{H}^Q$, respectively, and let $\rho\in\ca{S}(\ca{H}^A\otm\ca{H}^Q)$ be a quantum state thereon.
Fix arbitrary $C<\log{d_A}-S(A|Q)_\rho$, $\epsilon,\xi>0$ and choose sufficiently large $n\in\mbb{N}$.
Let $U_{k,i}$ be unitaries on $\ca{H}^A$ that are chosen independently and randomly according to the Haar measure for each $i\in[n]$ and $k\in[2^{nC}]$, and $U_k:=\bigotimes_{i=1}^nU_{k,i}$.
Then, with probability no smaller than $1-\xi$, there exists a POVM $\{\Lambda_k\}_{k=1}^{2^{nC}}$ on $A^nQ^n$ and it holds that
\alg{
1-
\frac{1}{2^{nC}}\!\sum_{k=1}^{2^{nC}}\!{\rm Tr}[\Lambda_k(U_k^{A^n}\otm I^{Q^n})\rho^{\otm n}(U_k^{A^n}\otm I^{Q^n})^\dagger]
\leq
\epsilon.\!
}
\elmm

\blmm{noncunion2}
Let $S$ be a finite-dimensional quantum system and let $\Lambda_1,\cdots,\Lambda_J$ be any sequence of positive semidefinite operators on $\ca{H}^S$ such that $0\leq\Lambda_j\leq I$.
Let $M_j$ be a qubit system for each $1\leq j\leq J$ and define a linear operator $\Pi_{\Lambda_j}:\ca{H}^S\rightarrow\ca{H}^S\otm\ca{H}^{M_j}$ by
$
\Pi_{\Lambda_j}
:=
\Lambda_j\otm\ket{0}+\sqrt{\Lambda_j}\sqrt{I-\Lambda_j}\otm\ket{1}.
$
For any subnormalized state $\varrho\in\ca{S}(\ca{H}^S)$, it holds that
\alg{
{\rm Tr}[\varrho]
-
{\rm Tr}[\hat{\Pi}\varrho\hat{\Pi}^\dagger]
\leq
2\sqrt{\sum_{i=1}^J{\rm Tr}[(I-\Lambda_j)\varrho]},
}
where $\hat{\Pi}:=\Pi_{\Lambda_J}\cdots\Pi_{\Lambda_1}$.
In addition, we have
$
{\rm Tr}[\hat{\Pi}\varrho\hat{\Pi}^\dagger]
=
{\rm Tr}[\hat{\Lambda}\varrho]
$,
where
\alg{
\hat{\Lambda}:=\sum_{x_1,\cdots,x_J=0,1}\Lambda_{1}^{(x_1)}\cdots\Lambda_{J}^{(x_J)}\cdot\Lambda_{J}^{(x_J)}\cdots\Lambda_{1}^{(x_1)}
}
and
$
\Lambda_{j}^{(0)}=\Lambda_j
$,
$
\Lambda_{j}^{(1)}=\sqrt{\Lambda_j}\sqrt{I-\Lambda_j}
$.
\elmm

\noindent
A proof of \rLmm{trans1} will be provided in \rSec{PRFtrans}.
The former half of \rLmm{noncunion2} was proved in Section 3 of \cite{wilde2013sequential} based on the non-commutative union bound for projective measurements \cite{sen2011achieving}.
The latter half follows by a straightforward calculation.

To prove the achievability of the rate tuple satisfying the condition \req{TTTe}, we invoke the notion of successive decoding which has been used e.g.~in the achievability proof of the classical capacity of a classical-quantum multiple access channel \cite{winter2001capacity}:
The receiver first decodes $M_1$ by performing a measurement on $A_1^nV^n$.
The measurement does not much change the state of the whole system as long as the error probability in decoding $M_1$ is sufficiently small.
The receiver then performs $U_{1,m_1}^\dagger$ on $A_1^n$ to reverse the encoding operation by Sender $1$.
In the second step, the receiver decodes $M_2$ by performing a measurement on $A_1^nA_2^nV^n$.
Since the encoding operation on $A_1^n$ has already been cancelled, in this step, $A_1^nV^n$ plays the same role as that of $V^n$ in the first step.
This time, the receiver then performs $U_{2,m_2}^\dagger$ on $A_2^n$ to reverse the encoding operation by Sender $2$.
Repeating this procedure $Z$ times, the receiver can decode all of the messages $M_1,\cdots,M_Z$ within a small error.

To be more precise, fix arbitrary $\epsilon>0$, arbitrary rate tuple $(C_1,\cdots,C_Z)$ satisfying the condition \req{TTTe}
and choose sufficiently large $n$.
For each $z\in[Z]$, we apply \rLmm{trans1} under the following correspondence:
\alg{
A\rightarrow A_z,\quad
Q\rightarrow A_1\cdots A_{z-1}V.
}
Let $U_{z,k_z,i}$ be unitaries on $\ca{H}^{A_z}$ that are chosen independently and randomly according to the Haar measure for $i\in[n]$, $k_z\in2^{c_z}$ and $z\in[Z]$, and let $U_{z,k_z}:=\bigotimes_{i=1}^nU_{z,k_z,i}$.
Let $I^{A_{<z}^n}$ be the identity operator on $(\bigotimes_{z'=1}^{z-1}\ca{H}^{A_{z'}})^{\otm n}$,
and define
\alg{
\!\!\rho_{z,k_z}\!:=\!(U_{z,k_z}\!\otm\! I^{A_{<z}^nV^n})(\rho^{A_{[z]}V})^{\otm n}(U_{z,k_z}\!\otm\! I^{A_{<z}^nV^n})^\dagger,
\laeq{moor}
}
It follows that, for any $z$ and with a probability no smaller than $1-\xi$, there exists a POVM $\{\Lambda_{z,k_z}\}_{k_z=1}^{2^{nC_z}}$ on $A_{[z]}^nV^n$ that satisfies
\alg{
1-\frac{1}{2^{nC_z}}\sum_{k_z=1}^{2^{nC_z}}\!{\rm Tr}[\Lambda_{z,k_z}\rho_{z,k_z}]
\leq
\epsilon.
\laeq{trans2}
}
Thus, with a probability no smaller than $(1-\xi)^Z\geq1-Z\xi$, there exist POVMs $\{\Lambda_{z,k_z}\}_{k_z=1}^{2^{nC_z}}$ satisfying \req{trans2} {\it for every} $z\in[Z]$.

We construct a decoding measurement $\{\Lambda_{{\bf k}}\}_{{\bf k}\in{\bf K}}$ from $\{U_{z,k_z}\}_{k_z=1}^{2^{nC_z}}$ and $\{\Lambda_{z,k_z}\}_{k_z=1}^{2^{nC_z}}$ as follows.
First, define
\alg{
\Upsilon_{z,k_z}
:=
(U_{z,k_z}\!\otm\! I^{A_{<z}^nV^n})^\dagger\!\sqrt{\Lambda_{z,k_z}}(U_{z,k_z}\!\otm\! I^{A_{<z}^nV^n}).\!
}
It follows that
\alg{
\!\!
(\Upsilon_{z,k_z})^2
\!:=\!
(U_{z,k_z}\!\otm\! I^{A_{<z}^nV^n})^\dagger\Lambda_{z,k_z}(U_{z,k_z}\!\otm\! I^{A_{<z}^nV^n}).\!\!
\laeq{trans1.5}
}
Thus, we have
\begin{eqnarray}
\sum_{k_z\in[K_z]}
(U_{z,k_z}\!\otm\! I^{A_{<z}^nV^n})(\Upsilon_{z,k_z})^2(U_{z,k_z}\!\otm\! I^{A_{<z}^nV^n})^\dagger
\nn\\
=
I^{A_{[z]}^nV^n}.
\;
\laeq{trans1.55}
\end{eqnarray}
Let $I^{A_{>z}^n}$ denote the identity operator on $(\bigotimes_{z'=z+1}^Z\ca{H}^{A_{z'}})^{\otm n}$.
From \req{moor} and \req{trans2}, we have
\alg{
1-\frac{1}{2^{nC_z}}\sum_{k_z=1}^{2^{nC_z}}\!{\rm Tr}[((\Upsilon_{z,k_z})^2\otm I^{A_{>z}})\rho^{\otm n}]
\leq
\epsilon.
\laeq{trans3}
}
Second, define
\alg{
\Upsilon_{z,k_z}^{(0)}:=\Upsilon_{z,k_z}^2,
\quad
\Upsilon_{z,k_z}^{(1)}:=\Upsilon_{z,k_z}\sqrt{I-\Upsilon_{z,k_z}^2}
}
for each $z$ and $k_z$.
It is straightforward to verify that
\alg{
(\Upsilon_{z,k_z}^{(0)})^2+(\Upsilon_{z,k_z}^{(1)})^2=\Upsilon_{z,k_z}^2.
\laeq{transT}
}
Third, we define
\alg{
\!
\Upsilon_{{\bf k}}^{x_1\cdots x_z}
\!:=\!
\Upsilon_{Z,k_Z}^{(x_Z)}(\Upsilon_{{Z-1},k_{Z-1}}^{(x_{Z-1})}\!\otm\! I^{A_Z})
\!\cdots\!(\Upsilon_{1,k_1}^{(x_1)}\!\otm\! I^{A_{>1}})
\!
\laeq{trans4}
}
for $x_1,\cdots,x_Z\in\{0,1\}$, where ${\bf k}=(k_1,\cdots,k_Z)$.
We apply \rLmm{noncunion2} under the correspondence
$
S\rightarrow A_{[Z]}^nV^n
$,
$
J\rightarrow Z
$,
$
j\rightarrow z
$,
$
\Lambda_j\rightarrow \Upsilon_{z,k_z}^2\otm I^{A_{>z}^n}
$
and
$
\varrho\rightarrow\rho^{\otm n}$.
It follows that
\alg{
&
1-
\sum_{x_1,\cdots,x_Z=0,1}{\rm Tr}[\Upsilon_{{\bf k}}^{x_1\cdots x_z}\rho^{\otm n}(\Upsilon_{{\bf k}}^{x_1\cdots x_z})^\dagger]
\nn\\
&\quad
\leq
2\sqrt{\sum_{z=1}^Z\left(1-{\rm Tr}[((\Upsilon_{z,k_z})^2\otm I^{A_{>z}})\rho^{\otm n}]\right)}.
}
Thus, from \req{trans3} and the concavity of the squareroot function, we have
\begin{eqnarray}
1-
\frac{1}{2^{nC_{[Z]}}}\sum_{{\bf k}\in{\bf K}}\sum_{x_1,\cdots,x_Z=0,1}{\rm Tr}[\Upsilon_{{\bf k}}^{x_1\cdots x_z}\rho^{\otm n}(\Upsilon_{{\bf k}}^{x_1\cdots x_z})^\dagger]
\nn\\
\leq
2\sqrt{Z\epsilon}.
\quad
\laeq{trans5}
\end{eqnarray}
Now, we construct a decoding measurement by
\alg{
\Lambda_{{\bf k}}
:=
\tilde{U}_{{\bf k}}\left(\sum_{x_1,\cdots,x_Z=0,1}(\Upsilon_{{\bf k}}^{x_1\cdots x_z})^\dagger\Upsilon_{{\bf k}}^{x_1\cdots x_z}\right)\tilde{U}_{{\bf k}}^\dagger,
}
where $\tilde{U}_{{\bf k}}:=(\bigotimes_{z=1}^ZU_{z,k_z})\otm I^{V^n}$.
From \req{trans1.55}, \req{transT} and \req{trans4}, it follows that
\alg{
\sum_{{\bf k}\in{\bf K}}\Lambda_{{\bf k}}=I,
}
which implies that $\{\Lambda_{{\bf k}}\}_{{\bf k}\in{\bf K}}$ is indeed a POVM.
Noting that $\rho_{\bf k}=\tilde{U}_{{\bf k}}\rho^{\otm n}\tilde{U}_{{\bf k}}^\dagger$, we also have
\alg{
{\rm Tr}[\Lambda_{{\bf k}}\rho_{{\bf k}}]
\!=
\!\sum_{x_1,\cdots,x_Z=0,1}\!\!\!{\rm Tr}[\Upsilon_{{\bf k}}^{x_1\cdots x_z}\rho^{\otm n}(\Upsilon_{{\bf k}}^{x_1\cdots x_z})^\dagger].\!
}
Substituting this to \req{trans5}, we arrive at
\alg{
1-\frac{1}{2^{nC_{[Z]}}}\sum_{{\bf k}\in{\bf K}}{\rm Tr}[\Lambda_{{\bf k}}\rho_{{\bf k}}]
\leq
2\sqrt{Z\epsilon}.
}
Since this relation holds for any small $\epsilon,\xi>0$ and any sufficiently large $n$, we complete the proof of the direct part.
\QED

\section{Proof of \rPrp{iindayo2}}
\lsec{DistRand}

In this section, we prove the distributed randomization theorem (\rPrp{iindayo2}).
We will use the same notations as in \rSec{DRThm}.

\subsection{Proof of The Converse Part}
Suppose that a rate tuple $(D_1,\cdots,D_Z)$ is achievable in distributed randomization of the state $\rho\in\ca{S}(\ca{H}^{A_1\cdots A_ZW})$.
Fix arbitrary $\vartheta>0$, choose sufficiently large $n$, and let $L_z=2^{nD_z}$ for each $z$.
By definition, there exists a set of unitaries $\mfk{U}_z\equiv\{U_{z,l_z}\}_{l_z=1}^{L_z}$ on $A_z^n$ for every $z$ such that
\alg{
\left\|\bar{\rho}^{A_{[Z]}^nW^n}
-
\pi^{A_{[Z]}^n}\otm(\rho^{\otm n})^{W^n}\right\|_1
\leq
\vartheta,
\!\!
\laeq{randCondII}
}
where 
\alg{
\bar{\rho}
:=
\left(\bigotimes_{z=1}^Z\ca{R}_z\right)(\rho^{\otimes n})
\laeq{randStateII}
}
and 
\alg{
\ca{R}_z(\cdot)
:=
\frac{1}{L_z}\sum_{l_z=1}^{L_z}U_{z,l_z}(\cdot)U_{z,l_z}^\dagger.
\laeq{dfnRkII}
}
Fix an arbitrary subset $\Gamma\subseteq[Z]$.
By tracing out $A_{\Gamma_c}^n$ in \req{randCondII}, we have
\alg{
\left\|\bar{\rho}^{A_{\Gamma}^nW^n}
-
\pi^{A_{\Gamma}^n}\otm(\rho^{\otm n})^{W^n}\right\|_1
\leq
\vartheta.
\!\!
\laeq{randCond2}
}

Define
$
{\bf L}_\Gamma:=\bigtimes_{z\in\Gamma}[L_z]
$.
Each element of ${\bf L}_\Gamma$ is denoted as ${\bf l}_\Gamma=(l_z)_{z\in\Gamma}$, where $l_z\in[L_z]$ for each $z$.
Correspondingly, define
$
U_{{\bf l}_\Gamma}
=
\bigotimes_{z\in\Gamma}U_{z,l_z}
$
and
\alg{
\rho_{{\bf l}_\Gamma}^{A_{\Gamma}^nW^n}
\!:=
(U_{{\bf l}_\Gamma}\otm I^{W^n})(\rho^{A_\Gamma W})^{\otm n}(U_{{\bf l}_\Gamma}\otm I^{W^n})^\dagger.
\laeq{titi1}
}
Let $L_\Gamma$ be a quantum system with a fixed orthonormal basis $\{|{\bf l}_\Gamma\rangle\}_{{\bf l}_\Gamma\in{\bf L}_\Gamma}$.
Consider a state
\alg{
\tilde{\rho}_\Gamma
:=
\frac{1}{|{\bf L}_\Gamma|}\sum_{{\bf l}_\Gamma\in{\bf L}_\Gamma}
\proj{{\bf l}_\Gamma}^{L_\Gamma}
\otm
\rho_{{\bf l}_\Gamma}^{A_{\Gamma}^nW^n}.
\laeq{titi2}
}
It is straightforward to verify that
\alg{
(\rho^W)^{\otm n}
&={\rm Tr}_{L_\Gamma A_{\Gamma}^n}[\tilde{\rho}_\Gamma],
\laeq{ramenn}
\\
\bar{\rho}^{A_{\Gamma}^nW^n}
&={\rm Tr}_{L_\Gamma}[\tilde{\rho}_\Gamma].
\laeq{ramen}
}

The entropies of $\tilde{\rho}_\Gamma$ is calculated as follows. We have
\alg{
&
\log{d_{L_\Gamma}}
\nn\\
&\geq
I(L_\Gamma:A_{\Gamma}^nW^n)_{\tilde{\rho}_\Gamma}
\laeq{frafraa}\\
&=
S(A_{\Gamma}^nW^n)_{\tilde{\rho}_\Gamma}
-
S(A_{\Gamma}^nW^n|L_\Gamma)_{\tilde{\rho}_\Gamma}
\laeq{frafrarara}\\
&=
S(W^n)_{\tilde{\rho}_\Gamma}
\!+\!
S(A_{\Gamma}^n|W^n)_{\tilde{\rho}_\Gamma}
\!-\!
S(A_{\Gamma}^nW^n|L_\Gamma)_{\tilde{\rho}_\Gamma}
\laeq{frafrara}\\
&=
nS(W)_\rho
\!+\!
S(A_{\Gamma}^n|W^n)_{\bar{\rho}}
\!-\!
S(A_{\Gamma}^nW^n|L_\Gamma)_{\tilde{\rho}_\Gamma},
\laeq{frafra}
}
where \req{frafraa} follows from the fact that $\tilde{\rho}_\Gamma$ is a classical-quantum state between $L_\Gamma$ and $A_\Gamma^nW^n$;
\req{frafrarara} from the definition of the mutual information;
\req{frafrara} due to the chain rule;
and \req{frafra} from \req{ramenn} and \req{ramen}.
Due to the condition \req{randCond2} and the Alicki-Fannes inequality \cite{alicki04,winter2016tight},  the second term in \req{frafra} is bounded as
\alg{
&
S(A_{\Gamma}^n|W^n)_{\bar{\rho}}
\\
&
\geq
S(A_{\Gamma}^n|W^n)_{\pi^{A_{\Gamma}^n}\otm\rho^{W^n}}
-2\vartheta\log{d_{A_{\Gamma}^n}}
\nn\\
&\quad\quad\quad-(1+\vartheta)h\left(\frac{\vartheta}{1+\vartheta}\right)
\\
&
=
n(1-2\vartheta)\log{d_{A_\Gamma}}-(1+\vartheta)h\left(\frac{\vartheta}{1+\vartheta}\right).
} 
Due to \req{titi2}, \req{titi1} and the unitary invariance of the von Neumann entropy, the third term in \req{frafra} is evaluated as
\alg{
&
S(A_{\Gamma}^nW^n|L_\Gamma)_{\tilde{\rho}_\Gamma}
=
\frac{1}{|{\bf L}_\Gamma|}\sum_{{\bf l}_\Gamma\in{\bf L}_\Gamma}
S\left(
\rho_{{\bf l}_\Gamma}^{A_{\Gamma}^nW^n}\right)
\\
&
=
\frac{1}{|{\bf L}_\Gamma|}\sum_{{\bf l}_\Gamma\in{\bf L}_\Gamma}
S\left(
(\rho^{A_\Gamma W})^{\otm n}\right)
=
nS(A_\Gamma W)_\rho.
}
In addition, we have
\alg{
\log{d_{A_\Gamma}}
=
\sum_{z\in\Gamma}\log{d_{A_z}}
}
and
\alg{
\log{d_{L_\Gamma}}
=
\sum_{z\in\Gamma}\log{d_{L_z}}
=
n\sum_{z\in\Gamma}D_z.
}
Combining the above relations, we obtain
\begin{eqnarray}
n\sum_{z\in\Gamma}D_z
\geq
n\left((1-2\vartheta)\sum_{z\in\Gamma}\log{d_{A_z}}
\!-\!
S(A_\Gamma|W)_\rho\right)\nn\\
-(1+\vartheta)h\left(\frac{\vartheta}{1+\vartheta}\right).\quad
\end{eqnarray}
Since this relation holds for any $\vartheta>0$ and sufficiently large $n$, we arrive at
\alg{
\sum_{z\in\Gamma}D_z
\geq
\sum_{z\in\Gamma}\log{d_{A_z}}
-
S(A_\Gamma|W)_\rho.
}
Noting that this relation holds for any $\Gamma\subseteq[Z]$, we complete the proof of the converse part.
\QED

\subsection{Proof of The Direct Part}

Recall that the region defined by \req{dontask} is a contrapolymatroid in $\bb{R}^Z$ and every extremal point $(D_1^*,\cdots,D_{[Z]}^*)$ is represented as
\alg{
\!\!
D_{\sigma(z)}^*
\!=\!
\log{d_{A_{\sigma(z)}}}
\!\!-\!
S(A_{\sigma(z)}|A_{\sigma(1)}\!\cdots\! A_{\sigma(z-1)}W)_{\rho},\!\!
}  
where $\sigma$ is a permutation on $[Z]$ (see \rLmm{polymatvertex}).  
Owing to the time-sharing scheme, it suffices to prove that all extremal points are in the achievable rate region.
Without loss of generality, it suffices to prove that a rate tuple $(D_1,\cdots,D_{[Z]})$ is achievable if
\alg{
D_{z}
>
\log{d_{A_{z}}}
-
S(A_{z}|A_{1}\cdots A_{z-1}W)_{\rho}
\laeq{TTTeII}
}  
for all $z\in[Z]$.
The proof is based on the following lemma:

\blmm{trans1D}
Let $A$ and $Q$ be finite-dimensional quantum systems represented by Hilbert spaces $\ca{H}^A$ and $\ca{H}^Q$, respectively, and let $\rho\in\ca{S}(\ca{H}^A\otm\ca{H}^Q)$ be a quantum state thereon.
Fix arbitrary $D>\log{d_A}-S(A|Q)_\rho$, $\vartheta,\xi>0$ and choose sufficiently large $n\in\mbb{N}$.
Let $U_{l,i}$ be unitaries on $\ca{H}^A$ that are chosen independently and randomly according to the Haar measure for each $i\in[n]$ and $l\in[2^{nD}]$.
Let $U_l:=\bigotimes_{i=1}^nU_{l,i}$ and $\ca{R}(\cdot):=\frac{1}{2^{nD}}\sum_{l=1}^{2^{nD}}U_l(\cdot)U_l^\dagger$.
Then, with probability no smaller than $1-\xi$, it holds that
\alg{
\left\|
\ca{R}(\rho^{\otm n})
-
\pi^{A^n}\otm(\rho^{\otm n})^{Q^n}
\right\|_1
\leq
\vartheta.
}
\elmm

\noindent
A proof of \rLmm{trans1D} will be provided in \rSec{PRFtrans}.

To prove the achievability of the rate tuple satisfying the condition \req{TTTeII}, we consider a successive protocol that proceeds as follows:
In the first step, a random unitary operation is applied on system $A_1^n$ so that the state on $A_1^n$ is completely randomized and is decorrelated from the remaining system $A_2^n\cdots A_Z^nE^n$.
Note that this operation does not affect the reduced state on $A_2^n\cdots A_Z^nE^n$.
In the second step, a random unitary operation is applied on $A_2^n$ to completely randomize $A_2^n$ and decorrelate it from $A_3^n\cdots A_Z^nE^n$.
Repeating this procedure $Z$ times, all subsystems $A_z^n$ are completely randomized and decoupled from each other and $W^n$.

To be more precise, fix arbitrary $\epsilon>0$, arbitrary rate tuple $(D_1,\cdots,D_Z)$ satisfying the condition \req{TTTeII}
and choose sufficiently large $n$.
For each $z\in[Z]$, we apply \rLmm{trans1D} under the following correspondence:
\alg{
A\rightarrow A_z,\quad
Q\rightarrow A_{z+1}\cdots A_{Z}W.
}
Let $U_{z,l_z,i}$ be unitaries on $\ca{H}^{A_z}$ that are chosen independently and randomly according to the Haar measure for each $i\in[n]$, $l_z\in[2^{nD_z}]$ and $z\in[Z]$.
Let $U_{z,l_z}:=\bigotimes_{i=1}^nU_{z,l_z,i}$ and $\ca{R}_z(\cdot):=\frac{1}{2^{nD_z}}\sum_{l=1}^{2^{nD_z}}U_{z,l_z}(\cdot)U_{z,l_z}^\dagger$.
It follows that, for each $z$ and with a probability no smaller than $1-\xi$, it holds that
\alg{
\left\|
\ca{R}_z((\rho^{A_z\cdots A_ZW})^{\otm n})-\pi^{A_z}\otm(\rho^{A_{z+1}\cdots A_ZW})^{\otm n}
\right\|_1
\leq
\vartheta.
\laeq{brubur}
}
Thus, with a probability no smaller than $(1-\xi)^Z\geq1-Z\xi$, the condition \req{brubur} holds {\it for every} $z\in[Z]$.
Let $A_{>z}$ denote the system $A_{z+1}\cdots A_{Z}$.
By the triangle inequality and the monotonicity of the trace distance, we have
\alg{
&
\!\left\|
\left(\bigotimes_{z=1}^Z\ca{R}_z\right)((\rho^{A_{[Z]}W})^{\otm n})-\pi^{A_{[Z]}}\otm(\rho^{W})^{\otm n}
\right\|_1
\nn\\
&\!\leq
\sum_{z=1}^Z
\left\|
\pi^{A_{[z-1]}}\otm\left(\bigotimes_{z'=z}^Z\ca{R}_{z'}\right)((\rho^{A_{\geq z}W})^{\otm n})
\right.
\nn\\
&\quad\quad\;
\left.
-
\pi^{A_{[z]}}\otm\left(\bigotimes_{z'={z+1}}^Z\ca{R}_{z'}\right)((\rho^{A_{\geq z+1}W})^{\otm n})
\right\|_1\!
\\
&
\!\leq
\sum_{z=1}^Z
\left\|
\ca{R}_z((\rho^{A_{\geq z}W})^{\otm n})\!-\!\pi^{A_z}\!\otm\!(\rho^{A_{\geq z+1}W})^{\otm n}
\right\|_1\!\!
\\
&
\leq 
Z\vartheta.
}
Since this relation holds for any small $\vartheta,\xi>0$ and any sufficiently large $n$,
we complete the proof of the direct part.
\QED

\section{Proof of \rLmm{trans1} and \rlmm{trans1D}}
\lsec{PRFtrans}

We prove \rLmm{trans1} and \rlmm{trans1D} based on the packing lemma and the covering lemma, respectively.
For the simplicity of presentations, we describe the two lemmas as a single one.
For the details, see Chapter 15 and 16 in \cite{wildetext}. 

\blmm{packing}
Consider an ensemble of states $\{p_x,\tau_x\}_{x\in\ca{X}}$ on a Hilbert space $\ca{H}$, and define
\alg{
\tau:=\sum_{x\in\ca{X}}p_x\tau_x.
}
Suppose there exists a projector $\Pi$ and a set of projectors $\{\Pi_x\}_{x\in\ca{X}}$ that satisfy
\alg{
\sum_{x\in\ca{X}}p_x{\rm Tr}[\Pi\tau_x]&\geq1-\varepsilon,
\\
\sum_{x\in\ca{X}}p_x{\rm Tr}[\Pi_x\tau_x]&\geq1-\varepsilon,
}
and there exist $\omega,\Omega,\omega',\Omega'>0$ such that
\alg{
{\rm Tr}[\Pi_x]\leq \omega,
\quad
\Pi\tau\Pi\leq\frac{1}{\Omega}\Pi
}
and
\alg{
{\rm Tr}[\Pi]\leq \Omega',
\quad
\Pi_x\tau_x\Pi_x\leq\frac{1}{\omega'}\Pi_x.
}
Let $\ca{M}$ and $\ca{M}'$ be finite sets, and $\ca{C}\equiv\{C_m\}_{m\in\ca{M}}$ and $\ca{C}'\equiv\{C_{m'}'\}_{m'\in\ca{M}'}$ be sets of random variables that take values in $\ca{X}$ independently according to a probability distribution $\{p_x\}_{x\in\ca{X}}$.
Then, 
\benum

\item There exists a POVM $\{\Lambda_m^{\ca{C}}\}_{m\in\ca{M}}$ for each $\ca{C}$ and satisfies
\alg{
&
\mbb{E}_{\ca{C}}
\left\{\frac{1}{|\ca{M}|}\sum_{m\in\ca{M}}{\rm Tr}[\Lambda_m^{\ca{C}}\tau_{C_m}]\right\}
\nn\\
&\quad
\geq
1-2(\varepsilon+2\sqrt{\varepsilon})-\frac{4\omega |\ca{M}|}{\Omega}.
}

\item
It holds that
\alg{
&
{\rm Pr}_{\ca{C}'}
\left\{
\left\|\bar{\tau}-\tau\right\|_1
\leq
\varepsilon+4\sqrt{\varepsilon}+24\sqrt[4]{\varepsilon}
\right\}
\nn\\
&\quad
\geq
1-2\Omega'\exp{\left(-\frac{\varepsilon^3}{4\ln{2}}\frac{|\ca{M}'|\omega'}{\Omega'}\right)},
}
where
$
\bar{\tau}:=|\ca{M}'|^{-1}\sum_{m'\in\ca{M}'}\tau_{C_{m'}'}
$.
\ennum
\elmm

\noindent
{\bf Proof of \rLmm{trans1} and \rlmm{trans1D}:}
Fix arbitrary $\delta,\varepsilon>0$ such that
\alg{
C+3\delta&\leq\log{d_A}-S(A|Q)_\rho,
\\
D-3\delta&\geq\log{d_A}-S(A|Q)_\rho
}
and
\alg{
3\varepsilon+4\sqrt{\varepsilon}
&<
\epsilon\xi,
\\
\varepsilon+4\sqrt{\varepsilon}+24\sqrt[4]{\varepsilon}
&<
\vartheta,
}
and choose sufficiently large $n$.
Let $\Pi_{n,\delta}^{Q^n}$ and $\Pi_{n,\delta}^{A^nQ^n}$ be projectors onto the $\delta$ typical subspaces of $(\ca{H}^Q)^{\otm n}$ and $(\ca{H}^{AQ})^{\otm n}$ with respect to $(\rho^Q)^{\otm n}$ and $(\rho^{AQ})^{\otm n}$, respectively.  
For each unitary $U$ on $(\ca{H}^A)^{\otm n}$ that is decomposed into $U=\bigotimes_{i=1}^nU_i$, define 
\alg{
\Pi_{n,\delta,U}^{A^nQ^n}&:=(U\otm I^{Q^n})\Pi_{n,\delta}^{A^nQ^n}(U\otm I^{Q^n})^\dagger,
\\
\rho_{n,U}&:=(U\otm I^{Q^n})\rho^{\otm n}(U\otm I^{Q^n})^\dagger.
}
It is straightforward to verify that
\alg{
\bar{\rho}_n
:=
\mbb{E}_U[\rho_{n,U}]
=
\pi^{A^n}\otm(\rho^Q)^{\otm n},
}
where the expectation is taken with respect to the Haar measure for each $U_i$.
We denote $I^{A^n}\otm\Pi_{n,\delta}^{Q^n}$ simply by $\tilde{\Pi}_{n,\delta}^{A^nQ^n}$.
Due to the property of the typical subspace, it holds that
\alg{
&\mbb{E}_U{\rm Tr}[\tilde{\Pi}_{n,\delta}^{A^nQ^n}\!\rho_{n,U}^{A^nQ^n}]
={\rm Tr}[\Pi_{n,\delta}^{Q^n}(\rho^Q)^{\otm n}]
\geq1-\varepsilon,\!
\nn\\
&\mbb{E}_U{\rm Tr}[\Pi_{n,\delta,U}^{A^nQ^n}\!\rho_{n,U}^{A^nQ^n}]
=
{\rm Tr}[\Pi_{n,\delta}^{A^nQ^n}\!\!(\rho^{AQ})^{\otm n}]
\geq1-\varepsilon.\!
\nn
}
In addition, we have
\alg{
&
{\rm Tr}[\Pi_{n,\delta,U}^{A^nQ^n}]
=
{\rm Tr}[\Pi_{n,\delta}^{A^nQ^n}]
\leq
2^{n(S(AQ)_\rho+\delta)},
\\
&
{\rm Tr}[\tilde{\Pi}_{n,\delta}^{A^nQ^n}]
\leq
2^{n(\log{d_A}+S(Q)_\rho+\delta)}
}
and
\alg{
&
\tilde{\Pi}_{n,\delta}^{A^nQ^n}\bar{\rho}_n\tilde{\Pi}_{n,\delta}^{A^nQ^n}
\leq 2^{-n(\log{d_A}+S(Q)_\rho-\delta)}\tilde{\Pi}_{n,\delta}^{A^nQ^n},
\\
&
\Pi_{n,\delta,U}^{A^nQ^n}\!\rho_{n,U}^{A^nQ^n}\!\Pi_{n,\delta,U}^{A^nQ^n}
\leq 2^{-n(S(AQ)_\rho-\delta)}\Pi_{n,\delta,U}^{A^nQ^n}.
}

We apply \rLmm{packing} under the following correspondence:
\alg{
\ca{X}\rightarrow\ca{U}((\ca{H}^A)^{\otm n}),&\quad p_x\rightarrow p(dU)
\\
\tau_x\rightarrow \rho_{n,U},&\quad \tau\rightarrow\bar{\rho}_n,
\\
\Pi\rightarrow \tilde{\Pi}_{n,\delta},&\quad \Pi_x\rightarrow\Pi_{n,\delta,U}.
}
We let
\alg{
\omega\rightarrow 2^{n(S(AQ)_\rho+\delta)},&\quad
\Omega\rightarrow 2^{n(\log{d_A}+S(Q)_\rho-\delta)},
\\
\ca{M}\rightarrow[2^{nC}],&\quad
\ca{C}\rightarrow\mfk{U}
}
and
\alg{
\omega'\rightarrow 2^{n(S(AQ)_\rho-\delta)},&\quad
\Omega'\rightarrow 2^{n(\log{d_A}+S(Q)_\rho+\delta)},
\\
\ca{M}'\rightarrow[2^{nD}],&\quad
\ca{C}'\rightarrow\mfk{U}.
}
It follows that there exists a POVM $\{\Lambda_k^{\mfk U}\}_{k=1}^{2^{nC}}$ on $A^nQ^n$ for each $\mfk{U}$ and it holds that
\alg{
&
\mbb{E}_{\mfk{U}}
\left\{\frac{1}{2^{nC}}\sum_{m=1}^{2^{nC}}{\rm Tr}[\Lambda_k^{\mfk U}U_k\rho^{\otm n}U_k^\dagger]\right\}
\nn\\
&\;
\geq
1-2(\varepsilon+2\sqrt{\varepsilon})-4\!\cdot\!2^{n(C-\log{d_A}+S(A|Q)_\rho+2\delta)}
\nn\\
&\;
\geq
1-2(\varepsilon+2\sqrt{\varepsilon})-4\!\cdot\!2^{-n\delta}.
}
By taking $n$ sufficiently large, and by applying Markov's inequality, we complete the proof of \rLmm{trans1}.
With $\ca{R}$ define as in \rLmm{trans1D}, we also have
\alg{
&
{\rm Pr}
\left\{
\left\|
\ca{R}(\rho^{\otm n})
-
\pi^{A^n}\otm(\rho^{\otm n})^{Q^n}
\right\|_1
\leq
\vartheta
\right\}
\nn\\
&
\geq
1\!-\!2\cdot 2^{n(\log{d_A}+S(Q)_\rho+\delta)}\exp{\left(\!-\frac{\varepsilon^3}{4\ln{2}}\!\cdot\!2^{n\delta}\right)}.\!
}
This yields \rLmm{trans1D} by taking $n$ sufficiently large.
\QED

\section{Conclusion and Discussion}
\lsec{cncl}

In this paper, we introduced the task of the quantum multiple-access one-time pad.
We considered an asymptotic limit of infinitely many copies and vanishingly small error, and derived a single-letter characterization of the achievable rate region.
Thereby we have provided a generalization of the quantum one-time pad \cite{schumacher2006quantum,brandao2012quantum} and the conditional quantum one-time pad \cite{sharma2020conditional} to a multi-sender scenario.
The proof of the converse part is based on a standard calculation of the entropies and the mutual informations, and that of the direct part is obtained by combining two subprotocols, i.e., distributed encoding and distributed randomization.
It is left open to obtain a similar characterization for the achievable rate region when the eavesdropper's side information is not necessarily a subsystem of the receiver's one.

A future direction is to extend the result to a one-shot scenario.
As in the case of the quantum multiple-access channels, successive decoding and time sharing may not be sufficient to derive the optimal one-shot rate region.
It would be necessary to employ the quantum joint typicality lemma \cite{sen2021unions} (see also the quantum multiparty packing lemma in \cite{ding2019quantum}), which has played a central role in the proof of the one-shot capacity theorems of the classical-quantum multiple-access channels \cite{sen2021unions,sen2021inner,chakraborty2021one,chakraborty2021oneinner}, and to address the simultaneous smoothing conjecture \cite{drescher2013simultaneous}, which has been one of the major open problems in quantum Shannon theory.

\section*{Acknowledgement}

The author thanks Atsushi Shimbo, Akihito Soeda and Mio Murao for useful discussions about local encoding.

\bibliographystyle{IEEEtran}
\bibliography{/Users/eyuriwakakuwa/Dropbox/DropTop/latexfiles/bibbib.bib}



%

\end{document}